\documentclass[pra,aps,twocolumn,amsmath,amssymb,floatfix,pra,reprint,footinbib,superscriptaddress,longbibliography]{revtex4}
\usepackage{graphics}      
\usepackage{graphicx}      
\usepackage{longtable}     
\usepackage{url}           
\usepackage{bm}    
\usepackage{braket}        
\usepackage{xcolor}
\usepackage{float}  
\usepackage{natbib}
\usepackage{graphicx}
\usepackage{comment}
\usepackage{siunitx}
\usepackage{appendix}
\usepackage{amsmath,amssymb,epsfig,color}

\usepackage[most]{tcolorbox}

\newcommand{\nl}{\nonumber \\}
\newcommand{\be}{\begin{equation}}
\newcommand{\ee}{\end{equation}}
\newcommand{\bea}{\begin{eqnarray}}
\newcommand{\eea}{\end{eqnarray}}

\newcommand{\tb}[1]{\textbf{\textit{#1}}}

\begin{document}

\title {Engineering Arbitrary Hamiltonians in Phase Space}

\author{Lingzhen Guo}
\affiliation{Center for Joint Quantum Studies and Department of Physics, School of Science, Tianjin University, Tianjin 300072, China}
\affiliation{Max Planck Institute for the Science of Light, Staudtstrasse 2, 91058 Erlangen, Germany}

\author{Vittorio Peano}
\affiliation{Max Planck Institute for the Science of Light, Staudtstrasse 2, 91058 Erlangen, Germany}


\begin{abstract}
We introduce a general method to engineer arbitrary Hamiltonians in the Floquet phase space of a periodically driven oscillator, based on the non-commutative Fourier transformation (NcFT) technique. We establish the relationship between an arbitrary target Floquet Hamiltonian in phase space and the periodic driving potential in real space. We obtain analytical expressions for the driving potentials in real space that can generate novel Hamiltonians in phase space, e.g., rotational lattices and sharp-boundary well. Our
protocol can be realised in a range of experimental platforms for nonclassical states generation and bosonic quantum computation.  
\end{abstract}

\date{\today}

\maketitle




\section{Introduction} 
Generation of nonclassical bosonic states \cite{Gerry2004book,Strekalov2019springer,Kubala2015njp}, e.g., squeezed lights, Fock states and  Schr\"odinger's cat states, is important not only for fundamental studies of quantum mechanics but also for applications in quantum technologies \cite{Braunstein2005RMP,Pan2012RMP,Strekalov2019springer,Yan2021fr}. 
For example, bosonic states with discrete translational or rotational symmetries  in phase space \cite{Leghtas2013prl,Mirrahimi2014njp,Heeres2017nc,Rosenblum2018science,Fluhmann2019nature,Hu2019nature,Campagne-Ibarcq2020nature,Gertler2021nature} have been proposed to encode quantum information \cite{Cochrane1999pra,Gottesman2001pra,Travaglione2002pra,Michael2016prx,Albert2018pra,Arne2020prx}, paving the way for hardware efficient quantum error correction \cite{Tzitrin2020pra,Terhal2020iop,Joshi2021qst,weizhou2021fr}.  Bosonic code states can be prepared and stabilized against dissipation via a sequence of universal  gates, e.g. interleaved \textit{selective number-dependent arbitrary phase} (SNAP) gates  and displacement gates \cite{Krastanov2015pra,fosel2020arxiv,Kudra2022prxq}. 
A series of recent works \cite{Puri2019PRX,Rymarz2021prx,Conrad2021pra,xanda2023arxiv} have pointed to an alternative approach based on  Hamiltonian engineering. The passive control introduced in this approach can be leveraged to facilitate fault tolerant operations, e.g. by suppressing phase flip errors \cite{Puri2019PRX},   suppressing dynamically the coupling to the environment \cite{Conrad2021pra}, and accelerating state preparation of code words \cite{xanda2023arxiv}.

Another area of interest for Hamiltonian engineering is topology. Due to the non-commutative nature of phase space,   a quantum particle moving on a closed phase-space loop acquires a  geometric phase analogous to the Aharonov-Bohm phase for a particles in a magnetic field. As a consequence, a gapped lattice Hamiltonian in phase space can support non-trivial Chern numbers \cite{Zaslavskii1986ZETF,Zaslavsky1991,Leboeuf1990prl,Leboeuf1992chaos,Gottesman2001pra,Billam2009pra,Guo2013prl,Zhang2017pra,Liang2018njp,Lorch2019prr}. This is an appealing feature because in a system with a physical  boundary, it would lead to topologically robust edge transport. While it has been shown how to generate arbitrary  lattice potentials in phase-space \cite{Guo2022prb}, so far it was unclear how to combine such  a potential with a sharp phase-space confinement. 

It is well known that the stroboscopic dynamics of any periodically driven system can be described in terms of a time-independent \emph{Floquet Hamiltonian} $\hat{H}_F$ defined via
\bea\label{eq-Floquet_Ham}
\exp\Big(\frac{1}{i\lambda}\hat{H}_FT\Big)\equiv\hat{U}(T,0)=\mathcal{T}\exp\left[\frac{1}{i\lambda}\int_{0}^{T}\hat{H}(t) dt\right]. \ \ 
\eea
Here, $\hat{U}(T,0)$ is the time-evolution operator with $T$ the time-period of the system's time-dependent Hamiltonian $\hat{H}(t)$. In adddition, $\lambda$ is an effective dimensionless Planck constant, and  $\mathcal{T}$ is the time-ordering operator. 
Except for very few models, it is impossible to obtain a closed  form of the  Floquet Hamiltonian $\hat{H}_F$ from the time-dependent Hamiltonian $\hat{H}(t)$. Instead, one often evaluates the Floquet  Hamiltonian  relying on a  high-frequency expansion \cite{Rahav2003pra,Goldman2014prx,Polkovnikov2015AIP}, e.g. the Magnus expansion theory \cite{Casas2001NJP,Blanes2009PR}, the van Vleck degenerate perturbation theory \cite{Eckardt2015NJP} and the Brillouin-Wigner perturbation theory \cite{Mikami2016prb}. In this work, we focus on the inverse problem, that is, to find the time-dependent Hamiltonian $\hat{H}(t)$ that synthetizes a target Floquet   Hamiltonian $\hat{H}^{(T)}_F$. This is the realm of Floquet engineering which is a very developed and active field \cite{Bukov2015aip,Liang2018njp,Rudner2020nrp,Jangjan2020scirep,Jangjan2022prb}. Most of the work so far has focused on implementing  specific Floquet Hamiltonians of  interest. However, a systematic constructive method to solve the inverse Floquet  problem for a single quantum particle  is still missing. In this work, we provide such a method.

\section{Model and goal}

As a starting point, we consider a periodically  driven  oscillator  with lab-frame Hamiltonian 
\bea\label{eq-HtVx}
\hat{\mathcal{H}}(t)=
\frac{\omega_0}{2}(\hat{p}^2+\hat{x}^2)+\beta V(\hat{x},t).
\eea
Here,  $\omega_0$ is the oscillator natural frequency, $\beta$ is the amplitude of the nonlinear driving potential  $V(\hat{x},t)$ which has time-period $T_d$ and might contain also static terms. In order to introduce an effective dimensionless  Planck constant $\lambda$ \cite{Marthaler2006pra,Peano2012prl,Guo2013PRA}, the position $\hat{x}$, the momentum  $\hat{p}$ and  $\hat{\mathcal{H}}(t)$ have been  rescaled such that $[\hat{x},\hat{p}]=i\lambda$ and at the same time  the Schr\"odinger equation reads $i\lambda\dot{\psi}= \hat{\mathcal{H}}(t)\psi$. Parameter $\lambda$ measures the quantumness of our system and $\lambda\to 0$ corresponds to the classical limit.

The Floquet Hamiltonian to be designed has time-period $T=2\pi/\omega_0$ and is defined via $\hat{O}(t)=\exp [i \hat{a}^\dagger\hat{a} \omega_0t]$, where $\hat{a}=(\hat{x}+i\hat{p})/\sqrt{2\lambda}$ is the annihilation operator. In other words, $\hat{H}(t)$ in Eq.~(\ref{eq-Floquet_Ham}) is the rotating-frame Hamiltonian given by
$
\hat{H}(t)=\hat{O}(t)\hat{\mathcal{H}}(t)\hat{O}^\dagger(t)+i\lambda \dot{\hat{O}}(t)\hat{O}^\dagger(t)
$, which in our case reads 
\bea\label{eq-Ht}
\hat{H}(t)=
\beta V[\hat{x}\cos(\omega_0t)+\hat{p}\sin(\omega_0t),t].
\eea 
We enforce the time-periodicity, $\hat{H}(t)=\hat{H}(t+T)$, by setting $T=qT_d$ with $q\in\mathbb{N}\ge 1$, corresponding to a $q$-photon resonance.
Any detuning from the multiphoton resonance is formally incorporated in the driving potential $V(x,t)$. For weak nonlinearity, $\beta\ll\omega_0$, the evolution in the rotating frame is slow. Thus,  we are in the realm of application of the Floquet high-frequency expansions, here, with the small parameter $\beta/\omega_0$.
This allows us to approximate the Floquet Hamiltonian with the leading order of the Floquet-Magnus expansion corresponding to the rotating wave approximation (RWA),
\bea\label{eq-h0h1h3}
\lim_{\omega_0/\beta\to \infty} \hat{H}_F(\hat{x},\hat{p})&=&\frac{1}{T}\int_{0}^{T}dt\hat{H}(t).
\eea

Our goal is to engineer an arbitrary target Floquet Hamiltonian $\hat{H}^{(T)}_F$ in phase space by properly designing the driving potential $V(x,t)$ in real space. Up to leading order (RWA) we, thus, require that the right-hand side of Eq.~(\ref{eq-h0h1h3})  coincides with the target Hamiltonian  $\hat{H}^{(T)}_F$. The ensuing solution becomes exact in the high-frequency limit $\omega_0/\beta\to \infty$.

\section{N\lowercase{c}FT technique}

As a preliminary step towards deriving a suitable driving potential $V(x,t)$, we introduce a useful decomposition of the target Hamiltonian $\hat{H}^{(T)}_F$ in the form of a \emph{noncommutative Fourier transformation} (NcFT). This can be viewed as a variant of quantum distribution theory \cite{scully1997quantum}. 
We wish to decompose the target Hamiltonian  $\hat{H}^{(T)}_F$ as a sum of plane-wave operators
\bea\label{eq-HTxp-mt}
\hat{H}^{(T)}_F
=
\frac{\beta}{2\pi}\iint dk_x dk_pf_T(k_x,k_p)e^{i(k_x\hat{x}+k_p\hat{p})}.
\eea
It can be shown that the Fourier coefficients $f_T(k_x,k_p)$ are given by the  inverse transformation [see App.~\ref{app-I}]  
\bea\label{eq-fFT-mt}
f_T(k_x,k_p)=\frac{e^{\frac{\lambda}{4}(k^2_x+k^2_p)}}{2\pi\beta}\iint dxdp H^{(T)}_Q(x,p) e^{-i(k_xx+k_pp)},\ 
\eea
where the phase-space function $H^{(T)}_Q(x,p)$ is the equivalent of the Husimi Q-function, here, for a Hamiltonian instead of the density operator.
We remind that the $Q$-function of an operator evaluated at a phase space point $(x,p)$  is simply its expectation value in the corresponding coherent state,  $H^{(T)}_Q(x,p)=\langle \alpha|\hat{H}^{(T)}_F|\alpha\rangle$ with  $\hat{a}|\alpha\rangle=\alpha|\alpha\rangle$ and $\alpha=(x+ip)/\sqrt{2\lambda}$. The latter mean value can be calculated by normal ordering the target Hamiltonian $\hat{H}^{(T)}_F(\hat{a}^\dagger,\hat{a})$. We point out three important features of the Hamiltonian $Q$-function: (i) For fixed $\lambda$, the mapping between Floquet Hamiltonians and $Q$-functions is one-to-one; (ii) The  Hamiltonian $Q$-function has the same  phase-space symmetries as the corresponding Floquet Hamiltonian, and (iii) a well-defined classical limit  $H^{(T)}_F(x,p)\equiv\lim_{\lambda\to 0} H^{(T)}_Q(x,p)$ [see App.~\ref{app-II} and App.~\ref{app-IX}].

\begin{figure}
\centerline{\includegraphics[width=\linewidth]{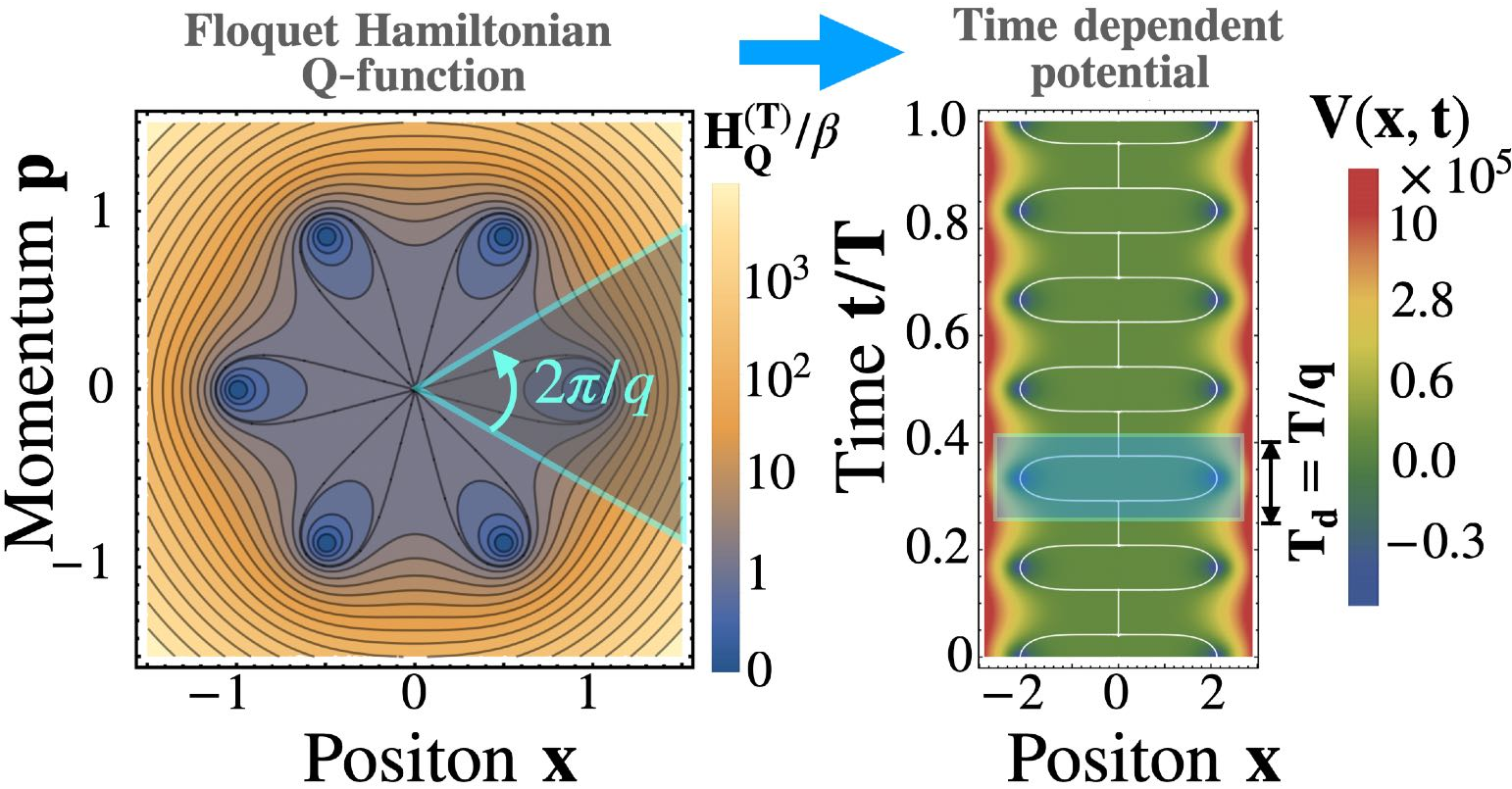}}
\caption{\label{Fig-RotationDegen}
{\bf Rotational lattice Hamiltonian in phase space.}
(Left) Q-function $H^{(T)}_Q(x,p)$ of target Floquet Hamiltonian Eq.~(\ref{eq-Han}); (Right) the engineered real-space potential $V(x,t)$
 for parameters $q=6$ and $\lambda=0.01$. The white contours indicate the minima of the instantaneous real-space potential. 
 }
\end{figure}

\section{Designing driving potential}

The  driving potential $V(x,t)$ that generates the target Floquet Hamiltonian $H^{(T)}_F(\hat{x},\hat{p})$ can be readily obtained  from its Fourier coefficient $f_T(k_x,k_p)$. We can formally write the solution as a  superposition of  sinusoidal potentials
\bea\label{eq-Vxt-1_mn}
 V(x,t)&=&\int_{0}^{+\infty}A(k,\omega_0 t)\cos[kx+\phi(k,\omega_0 t)]dk
\eea 
with time-varying amplitudes $A(k,\tau)$ and phases $\phi(k,\tau)$ determined from the Fourier coefficients in polar coordinates ($k_x=k\cos\tau$, $k_p=k\sin\tau$)
\bea\label{eq:ampl_and_phase}
A=k|f_T(k\cos \tau,k\sin \tau)|,\ \phi=\mathrm{Arg}f_T(k\cos \tau,k\sin \tau).\ 
\eea
This solution can be readily verified by plugging it into  Eqs.~(\ref{eq-Ht}) and (\ref{eq-h0h1h3}), and changing the integration variables back to cartesian coordinates to arrive at Eq.~(\ref{eq-HTxp-mt}) [see App.~\ref{app-III}]. 
In the remainder of this paper, we demonstrate the flexibility of our method by calculating the potential $V(x,t)$ for a range of interesting  Floquet Hamiltonians. In passing, we will also highlight more general features of our solution and comment on certain associated subtleties.

\section{Examples}

\subsection{Rotational  lattice} 
We now apply our method to engineer a  particularly interesting  Floquet Hamiltonian with $q$-fold symmetry in phase space  
\bea\label{eq-Han}
\hat{H}^{(T)}_F=\beta\left[(\hat{x}-i\hat{p})^q-1 \right]\left[(\hat{x}+i\hat{p})^q-1 \right].
\eea
The discrete rotational symmetry can be described by 
$\hat{R}(\frac{2\pi}{q})\hat{H}^{(T)}_F\hat{R}^{\dagger}(\frac{2\pi}{q})=\hat{H}^{(T)}_F$, where $\hat{R}(\theta)=\exp(i \hat{a}^\dagger\hat{a} \theta)$ is a phase-space rotation by an angle $\theta$ \cite{Guo2013prl,Arne2020prx}. This Hamiltonian supports $q$ global  minima, cf. the $Q$-function in Fig.~\ref{Fig-RotationDegen} (left) for $q=6$. Here, we have rescaled the phase-space coordinates such that the $q$ global minima fulfill $|x+ip|=1$ corresponding to different classical solutions. Remarkably, quantum fluctuations do not introduce any tunneling between these solutions as the corresponding coherent states $|\alpha_m\rangle=|e^{im\frac{2\pi}{q}}/\sqrt{2\lambda}\rangle$ with $m=0,1,\cdots,q-1$ are exact zero-energy eigenstates.  In other words,  the groundstate manifold is $q$-dimensional space spanned by $q$ $q$-legged cat states.

 Note that since the Hamiltonian $Q$-function  is a polynomial,  
its Fourier transform  Eq.~(\ref{eq-fFT-mt})  is divergent. To solve this problem, we renormalize the divergence introducing the bounded Hamiltonian $\hat{H}^{(T)}_{F\gamma}=U_\gamma\hat{H}^{(T)}_{F}U_\gamma$ with $U_\gamma\equiv e^{-\gamma \hat{a}^\dagger\hat{a}}$. Obviously, $\lim_{\gamma\to 0}\hat{H}^{(T)}_{F\gamma}=\hat{H}^{(T)}_{F}$. We can calculate  analytically $f_T(k_x,k_p)$ and $V(x,t)$ for $\hat{H}^{(T)}_{F\gamma}$ for any arbitrary positive integer $q$ and  $\gamma> 0$. 
This allows us to arrive at  a closed expression for the driving potential in the limit  $\gamma\to 0$ (see App.~\ref{app-IV})
\bea\label{eq:time-dep_rotational_sym_MT}
V(x,t)&=&\sum_{m=1}^{q}B_{q,m}\lambda^{q-m}x^{2m}-C_q\cos(q\omega_0 t)x^q,\ \ \ \ \ \ 
\eea
with $B_{q,m}=\frac{(2^mq!)^2(-1)^{q+m}}{(2m)!(q-m)!}$ and  $C_q=\frac{2\sqrt{\pi}q!}{\Gamma[(2q+3-(-1)^q)/4]}$. We note that for $q=2$ we recover a well-known result: Eq.~(\ref{eq:time-dep_rotational_sym_MT})  corresponds to a parametrically driven Duffing oscillator \cite{Marthaler2006pra,Peano2012prl,Dykman2012book,Dykman2022rmp}. We further note that the driving period is $T_d=T/q$ which directly follows from the $q$-fold rotational symmetry of the Floquet Hamiltonian. 

Realizing Hamiltonian (\ref{eq-Han}) is appealing in view of quantum computation because weak photon decay, with rate $\kappa\ll 2\beta q^2$, steers the oscillator towards its groundstate manifold containing code and error spaces of cat code  \cite{Mirrahimi2014njp}, see Ref.~\cite{Puri2019PRX} for $q=2$ and Appendix for the general case. In the App.~\ref{app-V}, we numerically verify the quality of the weak dissipation and rotating wave approximation for realistic parameters.

\begin{figure}
\centerline{\includegraphics[width=\linewidth]{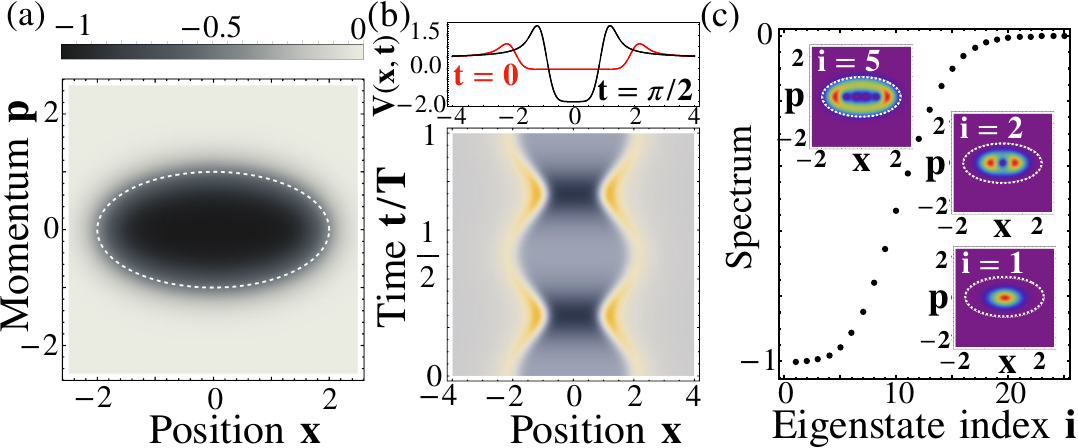}}
\caption{\label{Figures-EllipticalPotential}
{\bf Elliptical well in phase space.} {\bf (a)} Hamiltonian Q-function $H^{(T)}_Q(x,p)/\beta$ with long axis $a=2$, short axis $b=1$, $\lambda=0.1$ and convolution factor $\sigma=\sqrt{\lambda}$. {\bf (b)} Designed driving potential $V(x,t)$ in one Floquet period (lower) and at instants $t=0$, $t=\pi/2$ (upper).
{\bf (c) } Energy spectrum and Husimi Q-functions of ground state ($i=1$), first excited state ($i=2$), fourth ($i=5$) excited eigenstates. The dashed circles in (a) and (c) indicate the boundary of elliptical well.
}
\end{figure}

\subsection{Sharp-boundary well} 
Next, we demonstrate that our method allows us to engineer wells with a sharp boundary in phase space. For concreteness we choose an elliptical shape, i.e. $H^{(T)}_Q(x,p)=-\beta$ inside the white dashed line in  Fig.~\ref{Figures-EllipticalPotential}(a)   and $H^{(T)}_Q=0$ otherwise. 
In the classical limit $\lambda\to0$, our method allows us to find a closed-form solution for  $V(x,t)$ (see App.~\ref{app-VII}). However, our solution is divergent at two time-dependent positions. In addition, it does not directly apply to the quantum regime, $\lambda\neq 0$, because the dependence of  $V(x,t)$ on $\lambda$ is not analytical. This is due to the exponential factor in Eq.~(\ref{eq-fFT-mt}) leading to divergent NcFT coefficients $f_T(k_x,k_p)$  in the limit of large wavevectors, $k^2_x+k_p^2\to\infty$, for any $\lambda\neq 0$. We remove these unphysical features by smoothing out the target Floquet Hamiltonian by applying a convolution with a Gaussian kernel with standard deviation $\sigma$, cf. Fig.~\ref{Figures-EllipticalPotential}(a).  For $\sigma$ above a threshold, $\sigma>\sqrt{\lambda/2}$, the NcFT spectrum  $f_T(k_x,k_p)$ becomes integrable and, thus, leads to a  smooth solution for $V(x,t)$, cf. Fig.~\ref{Figures-EllipticalPotential}(b) and the closed expression in the App.~\ref{app-VII}.  This implies that we can implement a potential step that is arbitrarily sharp compared to the typical dimensions of phase-space well, but should remain smooth on the scale of the oscillator quantum fluctuations.  Note that Floquet Hamiltonians with sharper boundaries ($\sigma<\sqrt{\lambda/2}$) are well-defined but cannot be realized using our method  (see App.~\ref{app-II} and App.~\ref{app-VII}). The spectrum and first few eigenmodes are also shown in   Fig.~\ref{Figures-EllipticalPotential}(c). The latter are squeezed non-gaussian states.

\begin{figure}
\centerline{\includegraphics[width=0.95\linewidth]{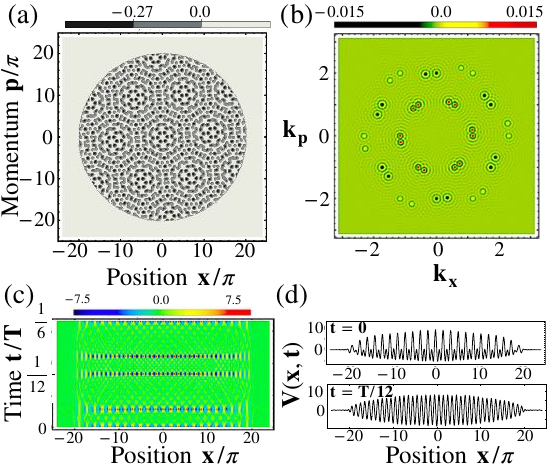}}
\caption{\label{Figures-MoirePotential}
{\bf Moir\'e superlattice  in phase space.} {\bf (a)} Hamiltonian Q-function of Moir\'e superlattice with twisted angle $\theta=10^\circ$ and confined in a region with radius $R=20\pi$. {\bf (b)} Density plot of NcFT coefficient $\beta f_T(k_x,k_p)$. {\bf (c)} Designed driving potential $V(x,t)$
for $ t\in [0,T/6)$. {\bf (d)} $V(x,t)$ at fixed time instants $t=0$ (upper) and $t=T/12$ (lower).
}
\end{figure}

\subsection{Moir\'e superlattice}
In Ref \cite{Guo2022prb}, we have shown how to synthesize arbitrary lattices in phase space. We can use our method to combine a lattice potential with a sharp confinement realizing a finite-size lattice. For concreteness we focus on  a Moir\'e superlattice, cf. Fig.~\ref{Figures-MoirePotential}(a). This is the phase-space equivalent of the 2D potential for electrons  in twisted graphene \cite{Rafi2011pnas,Cao2018nature-1,Cao2018nature-2,Thomson2018prb}.  The Moir\'e superlattice  is formed by overlaying two honeycomb  lattices  with a relative twist angle $\theta_0$ in a finite region of radius $R$. Outside of this region $H^{(T)}_F(x,p)=0$.  The resulting Hamiltonian Q-function for the twist angle $\theta_0=10^\circ$ is shown in Fig.~\ref{Figures-MoirePotential}(a). As discussed above, overall the Floquet Hamiltonian should be smooth on the scale of the oscillator quantum fluctuations. As for the phase-space well example above, this can be implemented by applying a convolution with a Gaussian kernel to the initially discontinuous Floquet Hamiltonian. The ensuing transition between the Floquet lattice potential and the phase-space region with $H^{(T)}_F(x,p)=0$   can be arbitrarily sharp compared to $R$ or to the honeycomb lattice constant. A closed formula for the Floquet Hamiltonian is given in the App.~\ref{app-VIII}.

Applying our method, we calculate the NcFT spectrum $f_T(k_x,k_p)$ shown in Fig.~\ref{Figures-MoirePotential}(b). It is formed by three groups of twelve peaks. Each group  of peaks is obtained from a single peak by applying one of the six-fold phase-space rotations and/or the rotation by the twist angle $\theta_0$, cf Fig.~\ref{Figures-MoirePotential}(a). The width of all the peaks is $\propto R^{-1}$. All these features as well as the exact locations of  the peaks can be read out from a closed-form solution for $f_T(k_x,k_p)$ given in the App.~\ref{app-VIII}.  
In Fig.~\ref{Figures-MoirePotential}(c), we plot the ensuing driving potential $V(x,t)$ for $0\leq t<T_d$. [In this case, the driving period $T_d$  is one-sixth of the natural period, $T_d=T/6$, reflecting the $6$-fold rotational-symmetry of our target Floquet Hamiltonian.]  In Fig.~\ref{Figures-MoirePotential}(d), we also plot the instant driving potential at $t=0$ and $ t=T_d/2$ (or $t=T/12$). We note that the real-space driving potential is a sequence of discrete lattice potentials localized in a finite region of real space that are switched on  for a short time interval. We note further that in the limit $R\to\infty$, the peaks in $(k_x,k_p)$-space become  $\delta$-functions, and the driving potential reduces to a discrete sequence of stroboscopic lattices  with  specific amplitudes, wavelengths, and phases  \cite{Guo2016pra,Liang2018njp,Guo2022prb,Guo2022book}. Considering that the contact interaction of cold atoms turns into a long-distance Coulomb-like interaction in the rotating frame \cite{Sacha2015pr,Sacha2017prb,Sacha2018prl,Liang2018njp,Sacha2020inbook,Guo2020njp,Guo2022prb,Guo2022book,Sacha2022aapps}, many atoms in the phase space Moir\'e superlattice   would mimic the behavior of electrons in twisted bilayer graphene \cite{Rafi2011pnas,Cao2018nature-1,Cao2018nature-2,Thomson2018prb}.

\subsection{ Artificial atomic spectrum}

\begin{figure}
\centerline{\includegraphics[width=\linewidth]{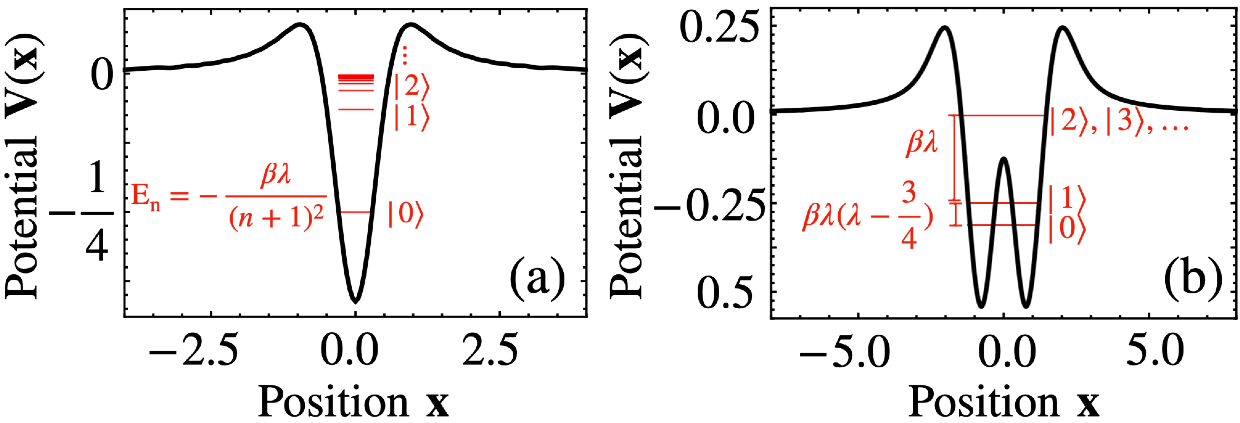}}
\caption{\label{Figures-FockPotential}
{\bf Artificial spectrum}:  {\bf (a)} designed hydrogen atomic levels with parameter $\lambda=\frac{1}{4}$; {\bf (b)} levels $|0\rangle$ and $|1\rangle$  gapped from other degenerate levels with $\lambda=1$. In both figures, the eigenstates $|n\rangle$ are the harmonic Fock states.
}
\end{figure}

Our method can be leveraged to implement a target spectrum $\{E_n\}$ as well as desired target eigenstates $\{|\psi_n\rangle\}$. As mentioned above, this could be useful for quantum simulations with interacting atoms.
In this scenario, our method could be straightforwardly applied to the target Floquet Hamiltonian
$
\hat{H}^{(T)}_F=\sum_{n}E_n|\psi_n\rangle\langle \psi_n|.
$ 
For concreteness, we consider $|\psi_n\rangle=|n\rangle$ where  $\{|n\rangle\}$  is the harmonic oscillator (Fock states) eigenbasis. In this example, the  Hamiltonian Q-function and the NcFT spectrum can be easily expressed as a sum over the excitation number $n$,
\bea\label{eq:Floq_ham_stat}
H^{(T)}_Q(x,p)=e^{-\frac{x^2+p^2}{2\lambda}}\sum_{n=0}^\infty\frac{E_n}{n!}\Big(\frac{x^2+p^2}{2\lambda}\Big)^n,
\eea
and
\bea\label{eq-FockfT}
f_T(k_x,k_p)=\sum_{n=0}^\infty\lambda\frac{E_n}{\beta} e^{\lambda\frac{ k^2}{4}} {_1F_1}(1+n;1;-\lambda\frac{ k^2}{2}),
\eea
respectively.
Here, ${_1F_1}(a;b;z)$ is the Kummer confluent hypergeometric function. The driving potential $V(x)$ can be straightforwardly calculated by plugging Eq.~(\ref{eq-FockfT}) into Eqs.~(\ref{eq-Vxt-1_mn}) and (\ref{eq:ampl_and_phase}).  Note  that since the NcFT spectrum  $f_T(k \cos\tau,k\sin\tau)$ is independent of the angular coordinate $\tau$, the driving potential $V(x)$ is \emph{static}. This, in turn, follows from our choice of eigenbasis leading to a target Floquet Hamiltonian invariant   under arbitrary phase-space rotations, cf. Eq.~(\ref{eq:Floq_ham_stat}).
Note further that the asymptotic behavior  ${_1F_1}(1+n;1;-\frac{ k^2}{2})\sim (-\frac{ \lambda k^2}{2n!})^ne^{-\frac{\lambda k^2}{2}}$ for $k\to\infty$ ensures that the integral in Eq.~(\ref{eq-Vxt-1_mn}) is well defined.
In Fig.~\ref{Figures-FockPotential} we display the potential $V(x)$ for two interesting  choices of the spectrum $\{E_n\}$. In panel (a), we fix $\{E_n\}$ to be the spectrum of the hydrogen atom $E_n=-{\beta\lambda}/{(n+1)^2}$. In panel (b) we choose $E_1=-\beta\lambda$ and $E_0=E_1-\beta\lambda(\lambda-\frac{3}{4})$
while all  other levels are zero, $E_{n> 1}=0$. Thus, at $\lambda=3/4$, the energies $E_0$ and $E_1$ of the second spectrum display an exact crossing. 

\section{State Preparation} 
Our method combined with an adiabatic ramp protocol following Ref.~\cite{xanda2023arxiv} can be exploited to prepare a desired quantum state. As example, we demonstrate the preparation of a cat state in the groundstate manifold of Hamiltonian  Eq.~(\ref{eq-Han}), including also the effects of dissipation,  see App.~\ref{app-VI}.

\section{Experimental implementations} 
In order to design arbitrary Hamiltonians in phase space, one needs the ability to engineer the driving real-space potential $V(x,t)$ in experiments.  This might be difficult in practice. An alternative route is to directly use   Eq.~(\ref{eq-Vxt-1_mn}). In the App.~\ref{app-VIB}, we show that the target Floquet Hamiltonian can be well-approximated  by replacing the integral with sum  of a finite number  of cosine lattice potentials. For example, we demonstrate the preparation of a three-legged cat state with $99\%$ fidelity using only 5 such potentials.
In  cold atom experiments, the building block  cosine lattice  is formed by laser beams intersecting at an angle \cite{Moritz2003prl,Hadzibabic2004prl,Guo2022prb}. In  experiments with superconducting circuits \cite{Chen2014prb,Hofheinz2011prl,Chen2011apl}, a microwave cavity in series with a Josephson junction (JJ) biased by a dc voltage ($V$) is described by the Hamiltonian $\hat{\mathcal{H}}(t)=\hbar\omega_0\hat{a}^\dagger\hat{a}-E_J\cos[\omega_Jt+\Delta(\hat{a}^\dagger+\hat{a})]$, where $E_J$ is the JJ energy, $\omega_J=2eV/\hbar$ is the Josephson frequency and $\Delta=\sqrt{2e^2/(\hbar \omega_0C)}$ with $C$ the cavity capacitance \cite{Armour2013prl,Gramich2013prl,Juha2013prl,Juha2015prl,Juha2016prb,Armour2015prb,Trif2015prb,Kubala2015iop,Hofer2016prb,Dambach2017njp,Lang2021njp,Lang2022arxiv}.  

\section{Summary and Outlook}
In this work, we have introduced a general constructive method to derive the driving potential, up to leading order in the Floquet-Magnus expansion, generating any arbitrary Floquet Hamiltonian  of a single Bosonic mode. We have also shown that,  in App.~\ref{app-V} and App.~\ref{app-VI}, it can be transferred to state-of-the-art experimental platforms to efficiently prepare  quantum states as part of a long-lived quantum memory. A natural extension of our work  would be  to include higher-order perturbative corrections as the inverse problem of the Floquet-Magnus theory. Another exciting prospect is to extend our method to a many-body scenario by upgrading  the single-particle plane-wave operator $\exp[{i(k_x\hat{x}+k_p\hat{p})}]$ used in   Eq.~(\ref{eq-HTxp-mt}) to a many-body equivalent $\exp[{\sum_ji(k^j_x\hat{x}_j+k^j_p\hat{p}_j})]$. In experiments with superconducting circuits, this could be implemented coupling a dc-voltage biased JJ  to multiple superconducting cavities \cite{Armour2013prl,Armour2015prb,Trif2015prb,Hofer2016prb,Dambach2017njp,Lang2022arxiv}.

\bigskip

\textbf{Acknowledgements}

 We acknowledge helpful discussions with Florian Marquardt and Muxin Han.

\appendix
 
\onecolumngrid



\section{~Noncommutative Fourier transformation}\label{app-I}

In this section, we provide detailed calculation of the noncomutative Fourier transformation (NcFT) coefficient for a given target Floquet Hamiltonian operator  $\hat{H}^{(T)}_F=\hat{H}^{(T)}_F(\hat{x},\hat{p})$. We start from writing  the target Hamiltonian as a sum of plane-wave operators, cf. Eq.~(5) in the main text,
\bea\label{eq-HTxp}
\hat{H}^{(T)}_F(\hat{x},\hat{p})
=
\frac{\beta}{2\pi}\int \int dk_x dk_pf_T(k_x,k_p)e^{i(k_x\hat{x}+k_p\hat{p})}.
\eea
In order to calculate the Fourier coefficient $f_T(k_x,k_p)$, we first express the target Hamiltonian with reordered ladder operators 
$$\hat{H}^{(T)}_F(\hat{a}^\dagger,\hat{a})\equiv\sum_{n,m}\chi_{nm} (\hat{a}^{\dagger})^n\hat{a}^m.$$ Note that the ordering here keeps all the terms from commutators. By defining the coherent state $|\alpha\rangle$ as the eigenstate of lowering operator $\hat{a}|\alpha\rangle=\alpha |\alpha\rangle$, we calculate the operator in the diagonal coherent representation 
\bea\label{eq-HTF-sm}
H^{(T)}_Q(\alpha,\alpha^*)\equiv\langle \alpha |\hat{H}^{(T)}_F|\alpha \rangle=\sum_{n,m}\chi_{nm}(\alpha^*)^n\alpha^m.
\eea
Function $H^{(T)}_Q(\alpha,\alpha^*)$ can also be written as $H^{(T)}_Q(x,p)$ by identifying $\alpha=(x+ip)/\sqrt{2\lambda}$ where
\begin{eqnarray}\label{}
\left\{
\begin{array}{lll}
x\equiv\langle\alpha|\hat{x}|\alpha\rangle=\sqrt{\frac{\lambda}{2}}(\alpha^*+\alpha)\\
p\equiv\langle
\alpha|\hat{p}|\alpha\rangle=i\sqrt{\frac{\lambda}{2}}(\alpha^*-\alpha).\\
\end{array}
\right.
\end{eqnarray}

In order to calculation the NcFT coefficient $f_T(k_x,k_p)$ in Eq.~(\ref{eq-HTxp}), we need to calculate the coherent diagonal element of the plane-wave operator $\langle \alpha|e^{i(k_x\hat{x}+k_p\hat{p})}|\alpha\rangle$. For this purpose, we introduce the displacement operator $\hat{D}_\alpha\equiv e^{\alpha
\hat{a}^\dagger-\alpha^*\hat{a}}$ with the following relationship \cite{Guo2020njp}
\begin{eqnarray}\label{eq-DD}
\hat{D}_\alpha \hat{D}_\beta=e^{i\mathrm{Im}(\alpha\beta^*)}\hat{D}_{\alpha+\beta},\ \ \hat{D}_\alpha
|\beta\rangle=e^{i\mathrm{Im}(\alpha\beta^*)}|\alpha+\beta\rangle.
\end{eqnarray}
We then write the plane-wave operator as $e^{i(k_x\hat{x}+k_p\hat{p})}=\hat{D}_{-\sqrt{\frac{\lambda}{2}}(k_p-ik_x)}$. Using the relationship (\ref{eq-DD}), we have the matrix element of plane-wave operator $e^{i(k_x\hat{x}+k_p\hat{p})}$ in coherent state representation
\begin{eqnarray}\label{eq-planewave}
\langle \alpha|e^{i(k_x\hat{x}+k_p\hat{p})}|\beta\rangle&=&\langle \alpha|\hat{D}_{-\sqrt{\frac{\lambda}{2}}(k_p-ik_x)}|\beta\rangle\nonumber\\
&=&\Big\langle
\alpha\Big|\beta-\sqrt{\frac{\lambda}{2}}(k_p-ik_x)\Big\rangle
e^{-i\sqrt{\frac{\lambda}{2}}\mathrm{Im}[(k_p-ik_x)\beta^*]}\nonumber\\
&=&e^{-\frac{1}{2}|\alpha|^2-\frac{1}{2}|\beta-\sqrt{\frac{\lambda}{2}}(k_p-ik_x)|^2
+\alpha^*\beta-\alpha^*\sqrt{\frac{\lambda}{2}}(k_p-ik_x)-i\sqrt{\frac{\lambda}{2}}\mathrm{Im}[(k_p-ik_x)\beta^*]} 
\end{eqnarray}
In the last step, we have used and the identity $\langle\alpha|\beta\rangle=e^{-|\alpha|^2/2-|\beta|^2/2+\alpha^*\beta}$.
Thus, we have the diagonal elements of plane-wave operator from Eq.~(\ref{eq-planewave})
\begin{eqnarray}\label{eq:diagH}
\langle \alpha|e^{i(k_x\hat{x}+k_p\hat{p})}|\alpha\rangle =\exp\big(-\frac{\lambda}{4}|k_p-ik_x|^2\big)e^{i(k_xx+k_pp)}=e^{-\frac{\lambda}{4}(k_x^2+k_p^2)}e^{i(k_xx+k_pp)}.
\end{eqnarray}
Using Eqs.~(\ref{eq-HTF-sm}) and (\ref{eq:diagH}),
we have the Fourier coefficient from Eq.~(\ref{eq-HTxp}) as follows
\bea\label{eq-fFT}
f_T(k_x,k_p)=\frac{e^{\frac{\lambda}{4}(k^2_x+k^2_p)}}{2\pi\beta}\int\int dxdp H^{(T)}_Q(x,p) e^{-i(k_xx+k_pp)}.\ 
\eea
Eqs.~(\ref{eq-HTxp}) and (\ref{eq-fFT}) construct the noncommutative Fourier transformation (NcFT) technique introduced in this paper.
From the hermicity of Hamiltonian operator, we have the following important relationship 
\bea\label{eq-H-fTkxkp-sm}
f_T(k_x,k_p)=f_T^*(-k_x,-k_p).
\eea
Note out that here we present a general way to calculation the NcFT coefficient. In practice, for some specific target Hamiltonians, there may exist a simpler and more direct way to obtain the result as for the rotational lattice shown below.

%

\section{~One-to-one correspondence between Hamiltonian operator and Q-function }\label{app-II}

In the above derivation of the NcFT coefficient for a given target Hamiltonian operator, we perform Fourier transformation of the Hamiltonian Q-function that only takes the diagonal elements of Hamiltonian operator in the coherent state representation, cf. Eq.~(\ref{eq-HTF-sm}). One may wonder if some information is lost by neglecting the off-diagonal elements. In this section, 
%
%
 we will prove the Hamiltonian operator $\hat{H}^{(T)}_F(\hat{x},\hat{p})$, given by Eqs.~(\ref{eq-HTxp}) and (\ref{eq-fFT}),
is fully determined by its Hamiltonian Q-function $H^{(T)}_Q(x,p)\equiv\langle \alpha |\hat{H}^{(T)}_F(\hat{x},\hat{p})|\alpha \rangle$ together with commutator $[\hat{x},\hat{p}]=i\lambda$. 

We write the Hamiltonian in the Fock representation $\hat{H}^{(T)}_F=\sum_{n,m}\xi_{nm}|n\rangle\langle m|$ with $n,m=0,1,\cdots$ and define the following operator
\bea
\hat{H}_{nm}(\hat{x},\hat{p})
&\equiv&
\frac{1}{2\pi}\int \int dk_x dk_pf_{nm}(k_x,k_p)e^{i(k_x\hat{x}+k_p\hat{p})}.
\eea
where $f_{nm}(k_x,k_p)$ is the NcFT coefficient of the operator $|n\rangle\langle m|$ given by
\bea
f_{nm}(k_x,k_p)&=&\frac{e^{\frac{\lambda}{4}(k^2_x+k^2_p)}}{2\pi}\int\int dxdp H_{nm}(x,p) e^{-i(k_xx+k_pp)} 
\eea
with the Q-function of operator $|n\rangle\langle m|$ given by
\bea
H_{nm}(x,p) =\langle \alpha|n\rangle\langle m|\alpha\rangle.
\eea
Because the target Hamiltonian is the liner superposition of $|n\rangle\langle m|$ with $n,m=0,1,\cdots$, we just need to prove 
\bea\label{eq-Hnm-sm}
\langle n'|H_{nm}(\hat{x},\hat{p})|m'\rangle=\delta_{n,n'}\delta_{m,m'}.
\eea
Using coherent state in the basis of Fock states $|\alpha\rangle=e^{-\frac{|\alpha|^2}{2}}\sum_n\frac{\alpha^n}{\sqrt{n!}}|n\rangle$, we calculate the Q-function of $|n\rangle\langle m|$
\bea
H_{nm}(x,p) &=&\langle \alpha|n\rangle\langle m|\alpha\rangle=\frac{1}{\sqrt{n!m!}}e^{-\frac{x^2+p^2}{2\lambda}}\Big(\frac{x-ip}{\sqrt{2\lambda}}\Big)^n\Big(\frac{x+ip}{\sqrt{2\lambda}}\Big)^m.
\eea
By introducing $x=r\cos\theta,\ p=r\sin\theta$ and $k_x=k\cos\tau,\ k_p=k\sin\tau$, we have the Fourier component
\bea\label{eq-sm-fnm}
f_{nm}(k,\tau)
&=&\frac{e^{\frac{\lambda}{4}k^2}}{2\pi}\int\int rdrd\theta H_{nm}(r\cos\theta,r\sin\theta) e^{-ikr\cos(\theta-\tau)}\nl
&=&\frac{e^{i(m-n)\tau}}{\sqrt{n!m!}}\Big(\frac{1}{\sqrt{2\lambda}}\Big)^{m+n}\frac{e^{\frac{\lambda}{4}k^2}}{2\pi}\int_0^{\infty}r^{m+n+1}e^{-\frac{r^2}{2\lambda}}dr\int_0^{2\pi}d\theta  e^{-ikr\cos(\theta-\tau)+i(m-n)(\theta-\tau)}\nl
&=&\frac{e^{i(m-n)\tau}}{\sqrt{n!m!}}\Big(\frac{1}{\sqrt{2\lambda}}\Big)^{m+n}e^{\frac{\lambda}{4}k^2}i^{n-m}\int_0^{\infty}r^{m+n+1}e^{-\frac{r^2}{2\lambda}}J_{n-m}(-kr)dr\nl
&=&e^{\frac{\lambda}{4}k^2}\sqrt{\frac{n!}{m!}}\Big(ie^{i\tau}\frac{1}{k}\sqrt{\frac{2}{\lambda}}\Big)^{m-n}\frac{\lambda}{\Gamma(1-m+n)}{_1F_1}(1+n;1-m+n;-\frac{\lambda}{2}k^2)
\eea
where $J_{n-m}(z)$ is the Bessel function with order of $n-m$, and ${_1F_1}(a;b;z)$ is the Kummer confluent hypergeometric function.
We introduce the marix element \cite{Guo2016njp} 
\bea\label{eq-sm-nem}
\langle n'| e^{i(k_x\hat{x}+k_p\hat{p})}|m'\rangle=e^{-\frac{\lambda}{4}(k^2_x+k^2_p)}\Big[\sqrt{\frac{\lambda}{2}}(k_p+ik_x)\Big]^{m'-n'}\sqrt{\frac{n'!}{m'!}}L_{n'}^{m'-n'}\big[\frac{\lambda}{2}(k^2_x+k^2_p)\big]
\eea
where $L_{n'}^{m'-n'}(z)$ is the generalized Laguerre polynomial. Then, we have the matrix element of $H_{nm}(\hat{x},\hat{p})$ in Fock representation
\bea
\langle n'| \hat{H}_{nm}(\hat{x},\hat{p})|m'\rangle&=&\frac{1}{2\pi}\int \int dk_x dk_pf_{nm}(k_x,k_p)\langle n'|e^{i(k_x\hat{x}+k_p\hat{p})}|m'\rangle\nl
&=&e^{i\frac{\pi}{2}(m'-n')}\sqrt{\frac{n'!}{m'!}}\sqrt{\frac{n!}{m!}}\Big(i\sqrt{\frac{2}{\lambda}}\Big)^{m-n}\frac{\lambda}{\Gamma(1-m+n)}\nl
&&\times\frac{1}{2\pi}\int_0^{2\pi}e^{i[(m-n)-(m'-n')]\tau} d\tau\nl
&&\times\int_0^{\infty}dk\Big(\sqrt{\frac{\lambda}{2}}k\Big)^{m'-n'}k^{n-m+1} {_1F_1}(1+n;1-m+n;-\frac{\lambda}{2}k^2)
L_{n'}^{m'-n'}\big(\frac{\lambda}{2}k^2\big)\nl
&=&\delta_{m-n,m'-n'}(-1)^{m-n}\sqrt{\frac{n'!}{(m-n+n')!}}\sqrt{\frac{n!}{m!}}\frac{\lambda}{\Gamma(1-m+n)}\nl
&&\times\int_0^{\infty}kdk \  {_1F_1}(1+n;1-m+n;-\frac{\lambda}{2}k^2)
L_{n'}^{m-n}\big(\frac{\lambda}{2}k^2\big)\nl
&=&\delta_{m-n,m'-n'}(-1)^{m-n}\sqrt{\frac{n'!}{(m-n+n')!}}\sqrt{\frac{n!}{m!}}\frac{2}{\Gamma(1-m+n)}\nl
&&\times\int_0^{\infty}\Tilde{k}d\Tilde{k} \  {_1F_1}(1+n;1-m+n;-\Tilde{k}^2)
L_{n'}^{m-n}\big(\Tilde{k}^2\big)   \ \ \ (\mathrm{where} \ \ \Tilde{k}=\sqrt{\frac{\lambda}{2}}k)\nl
&=&\delta_{n,n'}\delta_{m,m'}.
\eea
This is the identity (\ref{eq-Hnm-sm}) we aim to prove. As a result, for a fixed $\lambda$, the mapping between Floquet Hamiltonians and $Q$-functions is one to one.

Note that the exponentially suppression factor $e^{-\frac{\lambda}{4}(k^2_x+k^2_p)}$ in Eq.~(\ref{eq-sm-nem})  cancels the same exponentially increasing factor in Eq.~(\ref{eq-sm-fnm}). We will mention the sequence of this point in the discussion of sharp-boundary elliptical well below.

\section{~Designing driving potential} \label{app-III}

In this section, we show how to construct the driving potential from the NcFT coefficient such that its Floquet Hamiltonian equals to the target Hamiltonian in the leading order (RWA).
We introduce the polar coordinate system in $(k_x,k_p)$ space via ($k_x=k\cos\tau, k_p=k\sin\tau$), and write the Fourier expansion Eq.~(\ref{eq-HTxp}) as 
\bea\label{eq-Fktheta}
\hat{H}^{(T)}_F&=&\frac{1}{2\pi}\int_{0}^{2\pi}d\tau\int_{-\infty}^{+\infty}dk\frac{|k|}{2}f_T(k,\tau)e^{ik(\hat{x}\cos\tau+\hat{p}\sin\tau)}.\ \ \ \ 
\eea
Here, we have defined the Fourier component in the polar coordinate system via $f_T(k,\tau)\equiv f_T(k_x,k_p)$, and allow for negative $k$ via the relation $f_T(k,\tau)\equiv f^*_T(-k,\tau)$, cf. Eq.(\ref{eq-H-fTkxkp-sm}). 
From the Fourier component of the target Hamiltonian $f_T(k,\tau)$, we design the real space driving potential as follows
\bea\label{eq-Vxt-1}
V(x,t)&=&\int_{-\infty}^{+\infty}\frac{|k|}{2}f_T(k,\omega_0 t)e^{i kx}dk,
\eea
where the phase variable $\omega_0 t$ in time domain plays the role of angle $\tau$ in $(k_x, k_p)$ space.
By setting driving period $T_d=2\pi/(\omega_0 q)$, the Hamiltonian in the rotating frame, cf.  Eq.~(3) in the main text, becomes 
\bea\label{}
\hat{H}(t)&=&\int_{-\infty}^{+\infty}\frac{1}{2}|k|f_T(k, \omega_0t)e^{ik(\hat{x}\cos \omega_0t+\hat{p}\sin \omega_0t)}dk.\ \ \ 
\eea
By time averaging the above Hamiltonian, cf.  Eq.~(4) in the main text, and comparing the averaged result to Eq.~(\ref{eq-Fktheta}), one can directly find that the lowest-order Floquet-Magnus expansion (RWA)  gives the target Hamiltonian
$\hat{H}^{(T)}_F(\hat{x},\hat{p})$. 
We can also write the engineered driving potential in real space as $$V(x,t)=\int_{0}^{+\infty}|kf(k, \omega_0t)|\cos[kx+\phi(k,t)]dk,$$ where we have introduced phase $\phi(k,t)=\mathrm{Arg}[f_T(k,\omega_0t)]$ and used the property $f(-k, \omega_0t)=f^*(k, \omega_0t)$. Thus, the driving potential can be engineered by superposing a series of cosine lattice potentials with tunable amplitudes $|kf_T(k,\omega_0t)|$ and phases $\phi(k,t)$.

\section{~ Rotational lattice Hamiltonian}\label{app-IV}

We apply our method to engineer the target Floquet Hamiltonian with $q$-fold discrete rotational lattice symmetry in phase space 
%
\bea\label{eq-sm-HTq}
\hat{H}^{(T)}_{F\gamma}=\frac{\beta}{|\alpha_0|^{2q}}e^{-\gamma\hat{a}^\dagger\hat{a}}(\hat{a}^{\dagger q}-\alpha_0^{*q} )(\hat{a}^{q}-\alpha_0^{q} )e^{-\gamma\hat{a}^\dagger\hat{a}},
\eea
where the factor $e^{-\gamma\hat{a}^\dagger\hat{a}}$ with $\gamma>0$ is introduced to suppress the divergence of Hamiltonian in phase space. The above Hamiltonian is a generalised version of the rotational lattice Hamiltonian discussed in the main text, and it goes back to  Eq.~(9) by setting $\alpha_0=1/\sqrt{2\lambda}$.
Using the identity $$e^{-\gamma\hat{a}^\dagger\hat{a}}|\alpha\rangle=e^{-\frac{1}{2}(1-e^{-2\gamma})|\alpha|^2}|\alpha e^{-\gamma}\rangle,$$ we have the Hamiltonian Q-function as follows 
\bea\label{eq-sm-HTxp-q}
H^{(T)}_{Q\gamma}(x,p)&=&\langle \alpha|\hat{H}^{(T)}_{F\gamma}|\alpha\rangle\nl
&=&\frac{\beta}{|\alpha_0e^{\gamma}|^{2q}}\exp(-\frac{x^2+p^2}{2\lambda\sigma^2_\gamma})\Big|\Big(\frac{x+ip}{\sqrt{2\lambda}}\Big)^q-\alpha^q_0e^{q\gamma}\Big|^2.\ \ 
\eea
Here, we have defined the parameter $\sigma_\gamma=1/\sqrt{1-e^{-2\gamma}}$.
%

In order to obtain the analytical expression for the NcFT coefficient of Hamiltonian Q-function, we transform into the polar coordinate system by introducing $(x=r\cos\phi,p=r\sin\phi)$ and $(k_x=k\cos\tau,k_p=k\sin\tau)$. Plugging Eq.~(\ref{eq-sm-HTxp-q}) into Eq.~(\ref{eq-fFT}), we have
\bea\label{eq-fT-n}
f_T(k,\tau)&=&\frac{e^{\lambda k^2/4}}{2\pi\beta}\int_0^{+\infty}rdr\int_0^{2\pi}d\phi H^{(T)}_{Q\gamma}(r\cos\phi,r\sin\phi)e^{-ikr\cos(\phi-\tau)}\nl
&=&\frac{e^{\lambda k^2/4}}{|\alpha_0e^{\gamma}|^{2q}}\int_0^{+\infty}rdre^{-\frac{r^2}{2\lambda\sigma^2_\gamma}}\Big[\frac{r^{2q}}{(2\lambda)^q}J_0(kr)-\frac{(ir\alpha_0e^{\gamma-i\tau})^{q}}{(2\lambda)^{\frac{q}{2}}}J_{-q}(kr)-\frac{(-ir\alpha^*_0e^{\gamma+i\tau})^{q}}{(2\lambda)^{\frac{q}{2}}}J_{q}(kr)+|\alpha_0e^{\gamma}|^{2q}J_0(kr)\Big]\nl
&=&\frac{\lambda e^{\frac{\lambda}{4}k^2} \sigma^{2(q+1)}_\gamma}{|\alpha_0e^{\gamma}|^{2q}}\Big[n!{_1F_1}(1+q;1;-\frac{\lambda}{2}\sigma^2_\gamma k^2)-\Big(-ie^{-i\tau}\alpha_0e^{\gamma}\sqrt{\frac{\lambda}{2}}\Big)^qk^q{_1F_1}(1+q;1+q;-\frac{\lambda}{2}\sigma^2_\gamma k^2)\nl
&&-\Big(-ie^{i\tau}\alpha_0^*e^{\gamma}\sqrt{\frac{\lambda}{2}}\Big)^qk^q{_1F_1}(1+q;1+q;-\frac{\lambda}{2}\sigma^2_\gamma k^2)+\Big|\frac{\alpha_0e^{\gamma}}{\sigma_\gamma}\Big|^{2q}{_1F_1}(1;1;-\frac{\lambda}{2}\sigma^2_\gamma k^2)\Big].
\eea
Here, $J_q(\bullet)$ is the Bessel function of $q$-th order and  ${_1F_1}(a;b;\bullet)$ is the Kummer confluent hypergeometric function. 

In order to obtain an analytical expression for the driving potential $V(x,t)$, we introduce the following identities:\\
\bea
&\mathrm{Identity\  I:}\ &\int_{-\infty}^{+\infty}\frac{1}{2}|k|e^{\frac{\lambda}{4}k^2}k^q{_1F_1}(1+q;1+q;-\frac{\lambda}{2}\sigma^2_\gamma k^2)e^{ikx}dk\nl
&&=2^{q-1}(2\lambda\sigma^2_\gamma-\lambda)^{-\frac{q+2}{2}}\Big[q\Gamma(\frac{q}{2})(1+(-1)^q){_1F_1}(\frac{q+2}{2};\frac{1}{2};-\frac{x^2}{2\lambda\sigma^2_\gamma-\lambda})\nl
&&+4ix(1-(-1)^n)\Gamma(\frac{q+3}{2}){_1F_1}(\frac{q+3}{2};\frac{3}{2};-\frac{x^2}{2\lambda\sigma^2_\gamma-\lambda})\Big];\label{Id-1}
\eea
\bea
&\mathrm{Identity\  II:}\  &\int_{-\infty}^{+\infty}\frac{1}{2}|k|e^{\frac{\lambda}{4}k^2}{_1F_1}(1;1;-\frac{\lambda}{2}\sigma^2_\gamma k^2)e^{ikx}dk=\frac{2}{2\lambda\sigma^2_\gamma-\lambda}-\frac{4x}{(2\lambda\sigma^2_\gamma-\lambda)^{\frac{3}{2}}}D\big(\frac{x}{\sqrt{2\lambda\sigma^2_\gamma-\lambda}}\big);\label{Id-2}
\eea
\bea
&\mathrm{Identity\  III:}\  &\int_{-\infty}^{+\infty}\frac{1}{2}|k|e^{\frac{\lambda}{4}k^2}{_1F_1}(q+1;1;-\frac{\lambda}{2}\sigma^2_\gamma k^2)e^{ikx}dk\nl
&&=\int_{0}^{+\infty}ke^{\frac{\lambda}{4}k^2}{_1F_1}(q+1;1;-\frac{\lambda}{2}\sigma^2_\gamma k^2)\cos(kx)dk\nl
&&=\frac{1}{2}\sum_{m=0}^{+\infty}\frac{(-1)^m}{(2m)!}\int_{0}^{+\infty}e^{\frac{\lambda}{4}k^2}{_1F_1}(q+1;1;-\frac{\lambda}{2}\sigma^2_\gamma k^2)(kx)^{2m}dk^2\nl
&&=\frac{1}{2}\sum_{m=0}^{+\infty}\frac{(-1)^m}{(2m)!}x^{2m}\int_{0}^{+\infty}e^{\frac{\lambda}{4}z}{_1F_1}(q+1;1;-\frac{\lambda}{2}\sigma^2_\gamma z)z^mdz\ (\mathrm{here},\ z=k^2)\nl
&&=\frac{1}{2}\sum_{m=0}^{+\infty}\frac{(-1)^m}{(2m)!}x^{2m}\big(-\frac{4}{\lambda}\big)^{m+1}\Gamma(m+1){_2F_1}(m+1,q+1;1;2\sigma^2_\gamma).
\label{Id-3}
\eea
Here, $\Gamma(\bullet)$ is the Gamma function and $D(z)=e^{-z^2}\int_0^ze^{t^2}dt$ is the Dawson function. Identity Eq.~(\ref{Id-2}) is the special case of identity Eq.~(\ref{Id-1}) by setting $n=0$. In identity Eq.~(\ref{Id-3}), ${_2F_1}(a,b;c;z)$ is the hypergeometric function given by the path integral in the complex $\zeta$-plane \cite{SRIVASTAVA20121}
\bea
{_2F_1}(a,b;c;z)=\frac{\Gamma(c)}{\Gamma(a)\Gamma(b)}\frac{1}{2\pi i}\int_{-i\infty}^{+i\infty}\frac{\Gamma(a+\zeta)\Gamma(b+\zeta)\Gamma(-\zeta)}{\Gamma(c+\zeta)}(-z)^{\zeta}d\zeta.
\eea
The above integral is valid for $|\arg(-z)|\leq \pi-\epsilon$ ($0<\epsilon<\pi$) and $a,b\notin \mathbb{Z}^-_0$. For $|z|<1$,  the hypergeometric function can be written by the power series
\bea\label{2F1-sm-1}
{_2F_1}(a,b;c;z)=\sum_{k=0}^{\infty}\frac{(a)_k(b)_k}{(c)_k}\frac{z^k}{k!}, \ \ \ |z|<1
\eea
 with the Pochhammer symbol $(x)_k=\Gamma(x+k)/\Gamma(x)$. The analytical continuation of ${_2F_1}(a,b;c;z)$ into the domain $|z|>1$ can be realised from the following relationship \cite{SRIVASTAVA20121}
 \bea\label{2F1-sm-2}
 {_2F_1}(a,b;c;z)&=&\frac{\Gamma(c)\Gamma(b-a)}{\Gamma(b)\Gamma(c-a)}\big(-\frac{1}{z}\big)^a {_2F_1}(a,1-c+a;1-b+a;\frac{1}{z})\nl
 &&+\frac{\Gamma(c)\Gamma(a-b)}{\Gamma(a)\Gamma(c-b)}\big(-\frac{1}{z}\big)^b {_2F_1}(b,1-c+b;1-a+b;\frac{1}{z}).
 \eea
 The above relationship is valid for $|\arg(-z)|\leq \pi-\epsilon$ ($0<\epsilon<\pi$) and $a-b\notin \mathbb{Z}$. 
 From Eqs.~(\ref{eq-Vxt-1}) and (\ref{eq-fT-n})-(\ref{Id-3}), we obtain the analytical expression for the designed driving potential for finite values of  $\gamma>0$  and arbitrary complex number $\alpha_0$ as follows
 \bea\label{eq-Vgamma-sm}
 V_{\gamma}(x,t)&=&-\frac{(\lambda \sigma^2_\gamma)^{q+1}\Gamma(q+1)}{|\sqrt{\lambda}\alpha_0e^{\gamma}|^{2q}}\sum_{m=0}^{+\infty}\frac{\Gamma(m+1)}{\Gamma(2m+1)}\frac{x^{2m}}{2}\big(\frac{4}{\lambda}\big)^{m+1}{_2F_1}(m+1,q+1;1;2\sigma^2_\gamma)\nl
 &&-\frac{(\lambda \sigma^2_\gamma)^{q+1}2^{q-1}}{|\sqrt{\lambda}\alpha_0e^{\gamma}|^{2q}(2\lambda\sigma^2_\gamma-\lambda)^{\frac{q+3}{2}}}\Big[\Big(-ie^{\gamma-i\tau}\alpha_0\sqrt{\frac{\lambda}{2}}\Big)^q+\Big(-ie^{\gamma+i\tau}\alpha^*_0\sqrt{\frac{\lambda}{2}}\Big)^q\Big]\nl
 &&\times \Big[(1+(-1)^q)q\Gamma(\frac{q}{2})\sqrt{2\lambda\sigma^2_\gamma-\lambda}{_1F_1}(\frac{q+2}{2};\frac{1}{2};-\frac{x^2}{2\lambda\sigma^2_\gamma-\lambda})\nl
 &&+4ix(1-(-1)^q)\Gamma(\frac{q+3}{2}){_1F_1}(\frac{q+3}{2};\frac{3}{2};-\frac{x^2}{2\lambda\sigma^2_\gamma-\lambda})\Big]\nl
 &&+\frac{2\sigma^2_\gamma}{2\sigma^2_\gamma-1}-\frac{1}{\sqrt{\lambda}}\frac{4x\sigma^2_\gamma}{(2\sigma^2_\gamma-1)^{\frac{3}{2}}}D\big(\frac{x}{\sqrt{2\lambda\sigma^2_\gamma-\lambda}}\big).
 \eea
 
 Next, we discuss how to calculate driving potential $V_{\gamma}(x,t)$ in the limit of $\gamma\rightarrow 0$ ($\sigma_\gamma=1/\sqrt{1-e^{-2\gamma}}\rightarrow +\infty$). Using Eq.~(\ref{2F1-sm-1}), (\ref{2F1-sm-2}) and the Euler's reflection formula $\Gamma(z)\Gamma(1-z)=\pi/\sin(\pi z)$, we obtain the following series expansion of the hypergeometric function for $|z|>1$  and $q\in \mathbb{Z}^+_0$
  \bea\label{eq-2F1-sm-z}
 {_2F_1}(1+m,1+q;1;z)=\sum_{k=0}^{+\infty}\frac{\Gamma(m+k+1)^2(-1)^{m+q+1}}{\Gamma(1+m)\Gamma(1+q)\Gamma(m-q+k+1)}\frac{1}{k!}\big(\frac{1}{z}\big)^{k+m+1}.
 \eea
 Note that the parameter $m$ actually can take the whole real values, i.e., $m\in\mathbb{R}$. Although the above expansion is not defined for $m$ at negative integers $m\in \mathbb{Z}^-$, but the limit values do exist  and can be defined as the  values of the expansion for $m\in \mathbb{Z}^-$ . By plugging the above series expansion Eq.~(\ref{eq-2F1-sm-z}) and also the confluent hypergeometric function $ {_1F_1}(a;b;z)=\sum_{k=0}^{\infty}\frac{(a)_k}{(b)_k}\frac{z^k}{k!}$ into Eq.~(\ref{eq-Vgamma-sm}), we obtain the driving potential in the limit of $\gamma\rightarrow0$ ($\sigma_\gamma\rightarrow+\infty$)
 \bea\label{}
 V_{\gamma\rightarrow0}(x,t)&=&\frac{1}{|\sqrt{\lambda}\alpha_0|^{2q}}\sum_{m=0}^{+\infty}\sum_{k=0}^{+\infty}\frac{2^{2m-q}\lambda^{q-m}}{\Gamma(2m+1)}\frac{\Gamma(m+k+1)^2(-1)^{m+q}}{\Gamma(m-q+k+1)}\frac{x^{2m}}{k!}\big(\frac{1}{2 \sigma^2_\gamma}\big)^{k+m-q}\label{eq-Vxt-a}\\
 &&-\frac{1}{|\sqrt{\lambda}\alpha_0|^{2q}}\Big[\Big(-ie^{\gamma-i\tau}\alpha_0\sqrt{\frac{\lambda}{2}}\Big)^q+\Big(-ie^{\gamma+i\tau}\alpha^*_0\sqrt{\frac{\lambda}{2}}\Big)^q\Big]\label{eq-Vxt-d}\\
 &&\times \Big[(1+(-1)^q)\sum_{m=0}^{+\infty}\frac{\Gamma(1+m+\frac{q}{2})\Gamma(\frac{1}{2})}{\Gamma(\frac{1}{2}+m)}(-1)^m2^{\frac{q}{2}-m-1}(\lambda\sigma_\gamma^2)^{\frac{q}{2}-m}x^{2m}\label{eq-Vxt-b}\\
 &&+i(1-(-1)^q)\sum_{m=0}^{+\infty}\frac{\Gamma(\frac{3}{2}+m+\frac{q}{2})\Gamma(\frac{3}{2})}{\Gamma(\frac{3}{2}+m)}(-1)^m2^{\frac{q}{2}-m-\frac{1}{2}}(\lambda\sigma_\gamma^2)^{\frac{q}{2}-m-\frac{1}{2}}x^{2m+1}\Big]\label{eq-Vxt-c}\\
 &&+1.
 \eea
 In line~(\ref{eq-Vxt-a}), only terms that satisfy $k+m-q=0$ give nonzero contribution otherwise $\frac{1}{\Gamma(m+k-q+1)}\big(\frac{1}{2 \sigma^2_\gamma}\big)^{k+m-q}=0$ in the limit of $\sigma_\gamma=+\infty$ (note that Gamma function $\Gamma(z)=\infty$ for nonpositive integer argument $z\in\mathbb{Z}_0^-$).  In line~(\ref{eq-Vxt-b}), only terms with even integer $q$ and $m\leq q/2$ give nonzero contribution. Furthermore, we emphasise that the driving potential $V(x,t)$ is obtained from the RWA. In the rotating frame, the oscillating terms from $m<q/2$ cannot cancel the time-dependent parts given by terms that contain $e^{\pm i q\tau}$ in line~(\ref{eq-Vxt-d}). Therefore, the only nontrivial contribution comes from the term with $m=q/2$ in line~(\ref{eq-Vxt-b}). For the same reason, only nontrivial contribution comes from the term with $m=(q-1)/2$ in line~(\ref{eq-Vxt-c}). By neglecting these terms, we have the designed driving potential
\bea\label{}
 V(x,t)&=&\frac{1}{|\sqrt{2\lambda}\alpha_0|^{2q}}\sum_{m=0}^{q}(-1)^{m+q}\frac{(2^{m}q!)^2}{(2m)!(q-m)!}\lambda^{q-m}x^{2m}\nl
  &&-\frac{(-i)^q}{|\sqrt{2\lambda}\alpha_0|^{2q}}\Big[\Big(e^{-i\tau}\alpha_0\sqrt{2\lambda}\Big)^q+\Big(e^{i\tau}\alpha^*_0\sqrt{2\lambda}\Big)^q\Big]\nl
 &&\times \Big([1+(-1)^q]\frac{\Gamma(1+q)\Gamma(\frac{1}{2})}{\Gamma(\frac{1}{2}+\frac{q}{2})}\frac{(-1)^{\frac{q}{2}}}{2}+i[1-(-1)^q]\frac{\Gamma(q+1)\Gamma(\frac{3}{2})}{\Gamma(\frac{q}{2}+1)}(-1)^{\frac{q-1}{2}}\Big)x^{q}\nl
 &&+1.
 \eea
By taking the value of $\alpha_0=1/\sqrt{2\lambda}$, we obtain the driving potential given by Eq.~(10) shown in the main text, i.e.,
\bea\label{eq:time-dep_rotational_sym_SM}
V(x,t)&=&\sum_{m=0}^{q}\frac{(2^mq!)^2(-1)^{q+m}}{(2m)!(q-m)!}\lambda^{q-m}x^{2m}-\sqrt{\pi}q!\Bigg[\frac{1+(-1)^q}{\Gamma(\frac{q}{2}+\frac{1}{2})}+\frac{1-(-1)^q}{\Gamma(\frac{q}{2}+1)}\Bigg]\cos(q\tau)x^q+1\nl
&=&\sum_{m=0}^{q}B_{q,m}\lambda^{q-m}x^{2m}-C_q\cos(q\omega_0 t)x^q+1.\ \ \ \ \ \ 
\eea
Here, we have defined the coefficients $B_{q,m}=\frac{(2^mq!)^2(-1)^{q+m}}{(2m)!(q-m)!}$ and  $C_q=\sqrt{\pi}q!\Big[\frac{1+(-1)^q}{\Gamma(\frac{q}{2}+\frac{1}{2})}+\frac{1-(-1)^q}{\Gamma(\frac{q}{2}+1)}\Big]=\frac{2\sqrt{\pi}q!}{\Gamma[(2q+3-(-1)^q)/4]}$.


\begin{figure*}
\centerline{\includegraphics[width=\linewidth]{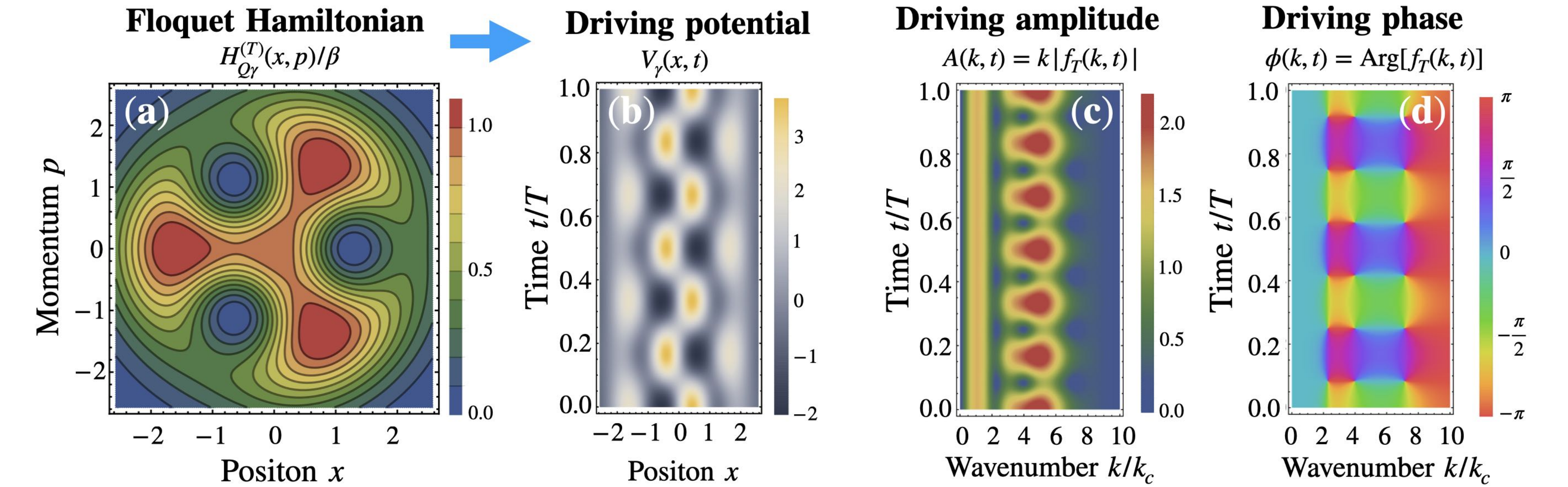}}
\caption{\label{Fig-sm-HT}
{\bf Target Hamiltonian and engineered driving potential.}
{\bf (a)} Q-function of target Hamiltonian $H^{(T)}_{Q\gamma}(x,p)/\beta$, cf. Eq.~(\ref{eq-sm-HTxp-q}), with $3$-fold rotational symmetry in phase space. {\bf (b)} The real-space driving potential $V_\gamma(x,t)$ that generates target Hamiltonian in (a). {\bf (c)-(d)} The time-varying amplitude $A(k,t)$ and phase $\phi(k,t)$ of cosine component of engineered driving potential $V_{\gamma}(x,t)$, cf. Eq.~(\ref{eq-sm-Vxt-1_mn}), as functions of wavenumber in the range $k\in (0,10k_c)$.  Parameters: $\lambda=1/4$, $\alpha_0=1/\sqrt{2\lambda}=\sqrt{2}$ and $\gamma=|\alpha_0|^{-2}/2=1/4$.
 }
\end{figure*}

\section{~Dissipative dynamics for the Bosonic code Floquet Hamiltonian}\label{app-V}
In this section, we discuss the dissipative dynamics for the Floquet Hamiltonian Eq.~(9) of the main text, which holds $q$-fold rotational symmetry in phase space. We prove that our arbitrary phase-space Hamiltonian engineering method can indeed allow us to design driving potential such that photon loss (dissipation) naturally leads the system state into the code subspace. Here, we focus on the experimentally relevant scenario occurring at a photon loss rate $\kappa$ that is much smaller than the typical frequency of the oscillations in the rotating frame (set by driving strength $\beta$). We mention in passing that the special case of $q=2$ (parametric oscillator) has been already discussed in Ref.~\cite{Puri2019PRX}. 

Note that the Floquet Hamiltonian $\hat{H}^{(T)}_F$ given by Eq.~(9) in the main text is the special case of the Floquet Hamiltonian $\hat{H}^{(T)}_{F\gamma}$ given by Eq.~(\ref{eq-sm-HTq}) in this Supplementary Material by taking parameters $\gamma=0$ and $\alpha_0=1/\sqrt{2\lambda}$. In fact, the Floquet Hamiltonian  $\hat{H}^{(T)}_F$ is somewhat ``unphysical" because it is divergent when the phase-space coordinates approach infinity. In contrast, the Floquet Hamiltonian $\hat{H}^{(T)}_{F\gamma}$ is more relevant to the real experiments and more convenient to be studied by numerical simulations since the divergence is suppressed by the exponential factor $e^{-\gamma\hat{a}^\dagger\hat{a}}$. For this reason, below, we discuss  the dissipative dynamics for the more general Floquet Hamiltonian $\hat{H}^{(T)}_{F\gamma}$. However, we anticipate that the same basic physics will apply to any sufficiently small $\gamma$ and, thus, also  to the limiting case $\gamma=0$, corresponding to the Floquet Hamiltonian $\hat{H}^{(T)}_F$ of the main text.

\subsection{~Groundstate manifold and engineered driving potential}

Before delving into the dissipative dynamics of the oscillator, we introduce a basis of rotationally invariant states for the ``ground-state'' manifold of Hamiltonian $\hat{H}^{(T)}_{F\gamma}$ and give more details about  the implementation of $\hat{H}^{(T)}_{F\gamma}$.
We remind that  the $q$-fold symmetry of the Hamiltonian $\hat{H}^{(T)}_{F\gamma}$ in phase space is described by 
\bea
\hat{R}_q\hat{H}^{(T)}_{F\gamma}\hat{R}_q^\dagger=\hat{H}^{(T)}_{F\gamma},
\eea
where $$\hat{R}_q\equiv e^{i\hat{a}^\dagger\hat{a}\frac{2\pi}{q}}$$ is the discrete rotational operator \cite{Guo2013prl,Arne2020prx}. 
According to the Bloch theorem extended in phase space \cite{Guo2013prl,Arne2020prx}, the eigentsates of $q$-fold rotational Hamiltonian can be written in form of 
\bea\label{eq-sm-psilm}
|\psi_{l,m}\rangle=\frac{1}{\sqrt{\mathcal{N}_{l,m}}}\sum_{p=0}^{q-1}e^{imp\frac{2\pi}{q} }\hat{R}_q^p|\phi_l\rangle,\ \ \ \ \ \hat{R}_q|\psi_{l,m}\rangle=e^{-im\frac{2\pi}{q}}|\psi_{l,m}\rangle
\eea
where index $l$ labels the Bloch bands,  $m$ is called quasinumber representing the \emph{parity} of state, $|\phi_l\rangle$ is the cell state of $l$-th Bloch band and $\mathcal{N}_{l,m}$ is the normalized factor. 
For the Hamiltonian $\hat{H}^{(T)}_{F\gamma}$ given by Eq.~(\ref{eq-sm-HTq}), the $q$ standard coherent states $|\alpha_0 e^{\gamma+i\frac{2\pi m}{q}}\rangle$ with $m=0,1,\cdots,q-1$ are the degenerate exact zero-energy eigenstates.  We thus choose the coherent state $|\alpha_0 e^{\gamma}\rangle$ as the cell state for the lowest band, $|\phi_0\rangle=\alpha_0 e^{\gamma}\rangle$,  and construct the Bloch states of lowest band 
\bea\label{eq-sm-psimcode}
|\psi_m\rangle=\frac{1}{\sqrt{\mathcal{N}_{m}}}\sum_{p=0}^{q-1}e^{imp\frac{2\pi}{q} }\hat{R}_q^p|\alpha_0e^{\gamma}\rangle=\frac{1}{\sqrt{\mathcal{N}_m}}\sum_{p=0}^{q-1}e^{im\frac{2\pi p}{q} }|\alpha_0 e^{\gamma+i\frac{2\pi p}{q}}\rangle.
\eea
Here, we have omitted the  band index $l=0$ for simplicity. The above $q$ $q$-legged cat states $|\psi_m\rangle$ construct the code subspace. 
For  example for $q=3$, the code space is panned by the following three code states 
\bea\label{eq-sm-3codestates}
|\psi_0\rangle&=&\frac{1}{\sqrt{\mathcal{N}_0}}\big(|\alpha_0 e^{\gamma}\rangle+|\alpha_0 e^{\gamma+i\frac{2\pi}{3}}\rangle+|\alpha_0 e^{\gamma+i\frac{4\pi}{3}}\rangle\big)\label{eq-sm-code0}\\
|\psi_1\rangle&=&\frac{1}{\sqrt{\mathcal{N}_1}}\big(|\alpha_0 e^{\gamma}\rangle+e^{i\frac{2\pi}{3} }|\alpha_0 e^{\gamma+i\frac{2\pi}{3}}\rangle+e^{i\frac{4\pi}{3} }|\alpha_0 e^{\gamma+i\frac{4\pi}{3}}\rangle\big)\label{eq-sm-code1}\\
|\psi_2\rangle&=&\frac{1}{\sqrt{\mathcal{N}_2}}\big(|\alpha_0 e^{\gamma}\rangle+e^{i\frac{4\pi}{3} }|\alpha_0 e^{\gamma+i\frac{2\pi}{3}}\rangle+e^{i\frac{2\pi}{3} }|\alpha_0 e^{\gamma+i\frac{4\pi}{3}}\rangle\big).\label{eq-sm-code2}
\eea

Next, we give more details about the implementation of $\hat{H}^{(T)}_{F\gamma}$. Also in this case we take as an example  for concrete numerical results the  case $q=3$, but our more qualitative discussion will apply equally well to other integer values of $q$. In Fig.~\ref{Fig-sm-HT}(a), we calculate and plot the corresponding Hamiltonian Q-function $H^{(T)}_{Q\gamma}(x,p)/\beta$, cf. Eq.~(\ref{eq-sm-HTxp-q}), for given parameters. The analytical expression for the driving potential $V_{\gamma}(x,\omega_0t)$ that generates the target Hamiltonian $\hat{H}^{(T)}_{F\gamma}$ is given by Eq.~(\ref{eq-Vgamma-sm}).  In Fig.~\ref{Fig-sm-HT}(b), we plot $V_{\gamma}(x,\omega_0t)$ in one time period $T=2\pi/\omega_0$. Different from the power-law driving potential given by Eq.~(10) in the main text, the driving potential here for generating $\hat{H}^{(T)}_{F\gamma}$ is confined in a finite region in real space due to the suppression factor $\gamma$. 
As discussed in the main text,  the  driving potential $V_\gamma(x,\omega_0t)$ can be formally written as a  superposition of cosine potentials
\bea\label{eq-sm-Vxt-1_mn}
 V_\gamma(x,\omega_0t)&=&\int_{0}^{+\infty}A(k,\omega_0 t)\cos[kx+\phi(k,\omega_0 t)]dk.
\eea 
Here, the time-varying amplitudes $A(k,\omega_0 t)$ and phases $\phi(k,\omega_0 t)$ are determined from the non-commutative Fourier coefficients $f_T(k,\omega_0t)$ given by Eq.~(\ref{eq-fT-n}) in polar coordinates ($k_x=k\cos\omega_0 t$, $k_p=k\sin\omega_0 t$)
\bea\label{eq-sm-ampl_and_phase}
A(k,\omega_0t)=k|f_T(k\cos \omega_0 t,k\sin \omega_0 t)|,\ \ \ \ \  \phi(k,\omega_0t)=\mathrm{Arg}[f_T(k\cos \omega_0 t,k\sin \omega_0 t)].\ 
\eea
 In Fig.~\ref{Fig-sm-HT}(c) and (d), we plot the time-varying amplitude $A(k,\omega_0t)$ and the phase $\phi(k,\omega_0t)$ respectively as functions of time $0\leq t \leq 2\pi/\omega_0$ and wavenumber $0\leq k\leq 10k_c$, where we have defined characteristic wavenumber 
 \bea\label{eq-sm-kc}
 k_c\equiv\sqrt{\frac{1-e^{-2\gamma}}{4\lambda}}.
 \eea

\subsection{~Dissipative dynamics leading to the groundstate manifold}

Here, we go back to the main goal of this section, i.e. to study the dissipative dynamics of the oscillator when the Floquet Hamiltonian $\hat{H}^{(T)}_{F\gamma}$ is prepared using our method. 
In the presence of weak photon loss and pure dephasing, the dissipative dynamics of the density matrix $\rho(t)$ is described by the  Lindblad master equation \cite{louisell1973book}
\bea\label{eq-sm-rho}
\frac{d}{dt}\rho(t)=-\frac{i}{\lambda}[\hat{\mathcal{H}}(t),\rho(t)]+\kappa \Big(\hat{a}\rho\hat{a}^\dagger-\frac{1}{2}\{\hat{a}^\dagger\hat{a},\rho\}\Big)+\eta \Big(\hat{a}^\dagger\hat{a}\rho\hat{a}^\dagger\hat{a}-\frac{1}{2}\{(\hat{a}^\dagger\hat{a})^2,\rho\}\Big),
\eea
where $\{\hat{A},\hat{B}\}\equiv\hat{A}\hat{B}+\hat{B}\hat{A}$ is the anticommutator, $\kappa$ is the single-photon loss rate and $\eta$ is the dephasing rate.  Here, $\hat{\mathcal{H}}(t)$ is the full Hamiltonian 
\bea\label{eq-sm-Vramp-0}
\hat{\mathcal{H}}(t)=
\frac{\omega_0}{2}(\hat{p}^2+\hat{x}^2)+\beta V_\gamma(x,\omega_0 t),
\eea
  with the potential $V_\gamma(x,t)$ as derived using our method, cf Eq.~(\ref{eq-sm-Vxt-1_mn}).  We note in passing that the master equation (\ref{eq-sm-rho}) is valid for any weakly non-linear high-quality-factor oscillator \cite{louisell1973book}. In our work, these two conditions translate into $\beta\ll\omega_0$, and  $\kappa,\eta\ll\omega_0$, respectively.
  Below we show that in the realistic parameter regime
\be\label{eq:cond_no_leak}
\eta|\alpha_e|^2 \ll \kappa \ll 2q^2\beta\ll \omega_0,
\ee 
 the oscillator tends to relax into the groundstate manifold of the  Bosonic code Floquet Hamiltonian. We will  corroborate our analytical derivation  with numerical results obtained by directly solving Eq.~(\ref{eq-sm-Vramp-0}), cf Fig.~2.  
 %

For our initial analytical treatment, we consider the dissipative dynamics   when photon decay is the only decay channel (setting the dephasing rate $\eta=0$) and switch to the rotating frame 
arriving at the Lindblad master equation \cite{louisell1973book,gilles1994pra,Marthaler2006pra,dykman2007pre,Terhal2020iop,Mirrahimi2014njp,Michael2016prx,Puri2019PRX,shruti2020sa} 
\begin{equation}\label{eq:Lindblad_RWA}
\frac{d}{dt}\rho_{rwa}=
-\frac{i}{\lambda}[\hat{H}^{(T)}_{F\gamma},\rho_{rwa}]
+\kappa \Big(\hat{a}\rho_{rwa}\hat{a}^\dagger-\frac{1}{2}\{\hat{a}^\dagger\hat{a},\rho_{rwa}\}\Big).
\end{equation}
Going from the lab-frame master equation~(\ref{eq-sm-rho}) to the rotating-frame master equation~(\ref{eq:Lindblad_RWA}), we have further simplified the description by applying  the  RWA to the Hamiltonian in the rotating frame. This approximation \cite{Casas2001NJP,Blanes2009PR} is standard and is consistent with the assumption $\beta\ll\omega_0$ leading to Eq.~(47).
We recall that the Floquet Hamiltonian $\hat{H}^{(T)}_{F\gamma}$ comprises $q$ separate wells. In the limit $\lambda\ll 1$, the transition between different wells occurs on a very large time scale (exponentially large in $\lambda^{-1}$). 
This means the total process towards equilibrium via photon loss can be divided into two distinguished stages: a fast  process of time scale $\kappa^{-1}$ relaxing to local equilibrium point $(x_e,p_e)$ in each local well and a slow transition (tunneling) process between local wells. The total probability in the code space is governed by the first quench process because the second slow transition process only adjusts the distribution over the states inside the code space. This makes it convenient to first analyze the linearized dynamics within one well. Below we follow the general  treatment of Ref.~\cite{dykman2011pra}.

For concreteness, we consider the well about  $\alpha_e=(x_e+ip_e)/\sqrt{2\lambda}$, cf. the minima of the Hamiltonian in Fig.~\ref{Fig-sm-HT}(a). Up to leading order in $\delta \hat{a}\equiv \hat{a}-\alpha_e$, we can approximate the local rotating wave Hamiltonian as a harmonic oscillator
\begin{equation}
\hat{H}^{(T)}_{F\gamma,\mathrm{local}}\approx 2\lambda q^2\beta   \delta \hat{a}^\dagger \delta \hat{a}. 
\end{equation}
Thus, the oscillations about each well have frequency  $2 \beta q^2$, giving rise to the quantized spectrum $E_n=\lambda 2 \beta q^2(n+1/2)$.  We note that the broadening of the Floquet spectrum due to photon decay is of the order $\lambda\kappa$. If the broadening is much smaller than the typical level spacing, $\kappa\ll 2q^2\beta$, the dissipative dynamics is well approximated by a rate equation for the Floquet states \cite{Marthaler2006pra,dykman2011pra}
\begin{equation}\label{eq-sm-Pl}
\frac{d}{dt}P_l=-\kappa\sum_{l\neq l'}
( W_{l'l}P_l
- W_{ll'}P_{l'}).
\end{equation}
Here, $P_l=\langle \phi_l|\rho|\phi_l\rangle$ is the occupation of the local Floquet states $|\phi_l\rangle$, i.e., the cell state of $l$-th Bloch band defined in Eq.~(\ref{eq-sm-psilm}).    
The transition rates $ W_{ll'}$ according to Fermi golden rule are given by
\bea
W_{ll'}=|\langle \phi_l|\hat{a}|\phi_{l'}\rangle|^2.
\eea 
Because the local ground state for the quasienergy well is simply the standard coherent state $|\phi_0\rangle=|\alpha_e\rangle$ according to our target setting, the transition rates from a local ground to the excited states are exactly zero, i.e., $W_{l0}=0$ for $l>0$. Thus, photon decay induces transitions only towards the bottom of the well. Obviously, the same analysis applies to all $q$ quasienergy wells. Thus, any superposition of states each localized about different quasi-energy wells will   relax towards the groundstate manifold spanned by the coherent states
$|\alpha_m\rangle=|e^{im\frac{2\pi}{q}}\alpha_e\rangle$ with $m=0,\ldots q-1$. 

\begin{figure*}
\centerline{\includegraphics[width=0.8\linewidth]{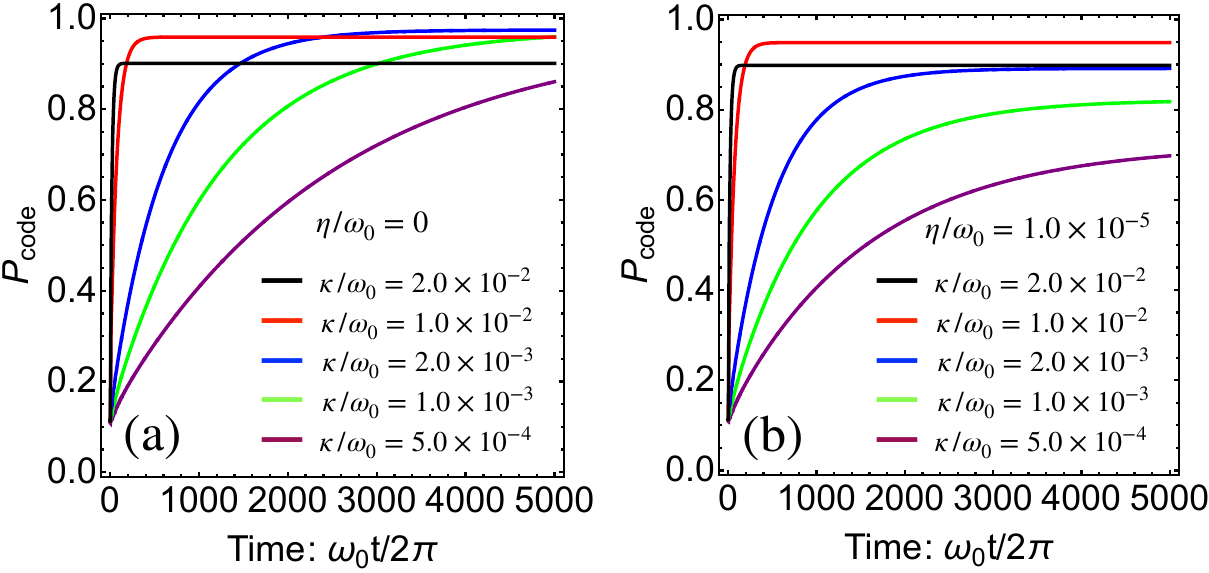}}
\caption{\label{Fig-SM-Dissipation}
{\bf Dissipative dynamics leading to code subspaces.} Time evolution of total probability $P_{code}$ over code space for different photon loss rates without pure dephasing (left) and with finite dephasing rate $\eta/\omega_0=10^{-5}$(right). Parameters: driving strength $\beta/\omega_0=1/40$ and driving frequency $\omega=\omega_0$.  
 }
\end{figure*}

We note in passing that Eq.~(\ref{eq-sm-Pl}) can be easily modified to account for finite photon dephasing with rate $\eta$ corresponding to the RWA Lindblad master equation
\cite{Marthaler2006pra,dykman2007pre,Terhal2020iop,Mirrahimi2014njp,Michael2016prx,Puri2019PRX,shruti2020sa}
\begin{equation}
\frac{d}{dt}\rho_{rwa}=
-\frac{i}{\lambda}[\hat{H}^{(T)}_{F\gamma},\rho_{rwa}]
+\kappa \Big(\hat{a}\rho_{rwa}\hat{a}^\dagger-\frac{1}{2}\{\hat{a}^\dagger\hat{a},\rho_{rwa}\}\Big)+\eta \Big(\hat{a}^\dagger\hat{a}\rho_{rwa}\hat{a}^\dagger\hat{a}-\frac{1}{2}\{(\hat{a}^\dagger\hat{a})^2,\rho_{rwa}\}\Big).
\end{equation}
In the limit $2q^2\beta\gg \kappa, \eta/(2\lambda)$, we arrive at Eq.~(\ref{eq-sm-Pl})
with  modified transition rates 
\bea
W_{ll'}=|\langle \phi_l|\hat{a}|\phi_{l'}\rangle|^2+\frac{\eta}{\kappa}|\langle \phi_l|\hat{a}^\dagger\hat{a}|\phi_{l'}\rangle|^2.
\eea 
 By making harmonic approximation near the bottom of  quasienergy well, the local Floquet levels are simple displaced Fock states $|\phi_l\rangle=\delta\hat{a}^{\dagger l}|\alpha_e\rangle/\sqrt{l!}$. The modified transition rates are approximately 
\bea 
 W_{ll'}\approx l'\left(1+\bar{n}_e\right)\delta_{l+1,l'}+(l'+1)\bar{n}_e\delta_{l-1,l'}
\eea 
leading  to a steady state  Boltzmann distribution over the quasienergy states with the effective thermal occupation number from rate equation (\ref{eq-sm-Pl})
\be
{\rm tr}(\rho \delta\hat{a}^\dagger \delta\hat{a})=\bar{n}_e=\frac{\eta}{\kappa}|\alpha_e|^2.
\ee
This corresponds to the steady state  groundstate manifold occupation probability 
\bea\label{eq-sm-P0}
P_0=\frac{1}{1+\bar{n}_e}=\frac{1}{1+\eta|\alpha_e|^2/\kappa}.
\eea
If the dephasing rate is weak enough compared to the photon loss rate satisfying $\eta|\alpha_e|^2/\kappa\ll 1$, the leakage probability is also small ($1-P_0\approx \eta|\alpha_e|^2/\kappa$). However, if the dephasing rate is strong so that $\eta|\alpha_e|^2/\kappa\gg1$, the leakage probability is large ($P_0\approx 0$) indicating that the state preparation via photon loss does not work anymore.

Summarizing our analysis so far, we can  conclude that  the oscillator  tends to relax into the groundstate manifold of the Bosonic code Floquet Hamiltonian  in the  parameter regime identified by the set of conditions shown in Eq.~(\ref{eq:cond_no_leak}).
The first condition ensures that  photon decay is the dominant decay channel, the second that the broadening of the Floquet levels is small, and the third that the RWA is valid. This is a realistic parameter regime in experiments with superconducting circuits.

This conclusion is also corroborated by numerical results, shown in Fig.~2, obtained by simulating the full dissipative dynamics as defined by the master equation (\ref{eq-sm-rho}), which includes   the full time-dependence of the coherent Hamiltonian Eq.~(\ref{eq-sm-Vramp-0}).
In these simulations, for the case $q=3$, we start from the zero-excitation Fock state and choose the driving strength $\beta=1/40$. 
In Fig.~\ref{Fig-SM-Dissipation}(left),  we show the time evolution of total probability of density matrix over the three-fold rotational code states given by Eq.~(\ref{eq-sm-3codestates}), i.e., 
\bea\label{eq-sm-Pcode}
P_{code}(t)\equiv \sum_{m=0}^{q-1}\langle\psi_m|\rho_{pre}(t)|\psi_m\rangle
\eea
for different photon loss rates at zero dephasing rate $\eta/\omega_0=0$. The probability $P_{code}(t)$ increases monotonously from an initial small value $P_{code}(0)\approx 0.1$ with faster speed for stronger photon loss. $P_{code}(t)$ is larger as the photon loss rate becomes stronger till an optimal value $\kappa/\omega_0=2.0\times 10^{-3}$ (up to $P_{code}>0.95$), then drops again as photon loss rate continues increasing.  This is because when the photon loss rate becomes large enough so that  the condition $\kappa\ll2\beta q^2\omega_0\approx 0.4\omega_0$ is not satisfied, the off-diagonal matrix elements of density matrix play the role and the validity of rate equation (\ref{eq-sm-Pl}) breaks down.
In Fig.~\ref{Fig-SM-Dissipation}(right), we show the time evolution of $P_{code}(t)$ for different photon loss rates with finite dephasing rate $\eta/\omega_0=10^{-5}$. The results are qualitatively the same as that in Fig.~\ref{Fig-SM-Dissipation}(left) but with a slightly optimal point of photon loss rate and a lower probability in code space due to the effective temperature from pure dephasing, cf. Eq.~(\ref{eq-sm-P0}).

\subsection{~Dissipative dynamics within the code and error spaces}

Next, we briefly discuss the dynamics within the groundstate manifold of the Floquet Hamiltonian (\ref{eq-sm-HTq}). As we explain below this manifold can be chosen to host the code and error spaces for a cat code. Here, we assume that the conditions in Eq.~(\ref{eq:cond_no_leak}) are fulfilled and go back to the simplified description without counter-rotating terms and dephasing. It is convenient to rewrite the Lindblad master equation projected onto the groundstate manifold  into a  basis of q-legged cat states, cf. Eq.~(\ref{eq-sm-psimcode}),
\bea\label{eq-cat-states}
|\psi_m\rangle=\frac{1}{\sqrt{\mathcal{N}_m}}\sum_{p=0}^{q-1}e^{im\frac{2\pi p}{q} }|\alpha_0 e^{\gamma+i\frac{2\pi p}{q}}\rangle.
\eea
Here, ${\cal N}_m$  are normalization constants and the quantum number $m$ can be interpreted as  the quasi-angularmomentum  or, equivalently, the photon number modolus $q$. We note that in the semi-classical limit, $\lambda\to 0$, all normalization converge exponentially fast to the same value, $\lim_{\lambda\to 0}{\cal N}_m= q$. We note further that the annihilation operator increases the value of the quasi-angularmomentum,
\be
\hat{a}|\psi_m\rangle=\alpha_0e^\gamma\sqrt{\frac{{\cal N}_{m+1}}{{\cal N}_{m}}}|\psi_{m+1}\rangle.
\ee
Thus, the annihilation operator $\hat{a}$ projected onto the groundstate manifold reads
\be
\hat{\Pi}_0\hat{a}\hat{\Pi}_0=\alpha_0e^\gamma\sum_{m=0}^{q-1} \sqrt{\frac{{\cal N}_{m+1}}{{\cal N}_{m}}}|\psi_{m+1}\rangle \langle \psi_{m}|\approx\alpha_0e^\gamma\sum_{m=0}^{q-1} |\psi_{m+1}\rangle \langle \psi_{m}|,
\ee
where $\Pi_0$ is the projector on the groundstate manifold. If $\lambda$ is small enough such that the last approximation is accurate, the dissipative dinamycs of density matrix $\rho_{\mathrm{code}}$ within the groundstate manifold is described by the simple master Lindblad Master equation
\be
\frac{d}{dt}\rho_{\mathrm{code}}=\kappa^\prime \Big(\hat{L}\rho\hat{L}^\dagger-\frac{1}{2}\{\hat{L}^\dagger\hat{L},\rho\}\Big)
\ee
 with the simple jump operator $\hat{L}=\sum_{m=0}^{q-1} |\psi_{m+1}\rangle \langle \psi_{m}|$ and damping rate  $\kappa'=\kappa|\alpha_0e^\gamma|^2$.
The groundstate manifold can be chosen to host the code and error spaces for a bosonic code if 
 $q=2(N+1)$. In this case, the code space is spanned by the cat states  with quasi-angular momentun $m=0$ and $m=N+1$. The remaining cat states span the error space. Such a code allows to correct the simultaneous decay of up to $N$ photons. This can be straightforwardly verified by applying the corresponding Knill-Laflamme conditions \cite{chuang2010book}.
 
 The standard cat code (correcting single photon decay) corresponds to the case $q=4$ (or $N=1$) \cite{Terhal2020iop}. In this case, the syndrome is a simple parity measurement  (code states have even parity) \cite{Schoelkopf2018science}. When every error is detected by applying repeated  parity measurements,  they can also be corrected by  updating the definition of the code and error states  using the mapping $m\to m+1$, e.g.  after the first  error is detected the odd (even) states become the code (error) states.

\section{~State preparation via adiabatic ramp }\label{app-VI}

In this section, we combine our method with the ``adiabatic ramp"  protocol recently introduced by X. C. Kolesnikow et al. in Ref.~\cite{xanda2023arxiv} to prepare a target rotational bosonic code state that is the ``ground state" of the Hamiltonian Eq.~(\ref{eq-sm-HTq}). We further discuss the complexity of implementing our driving scheme by superconducting circuit with Josephson junctions (JJs) or ultracold atom in optical lattice with beam lasers, and show that the quality of final prepared state is still very good even the number of  JJs or laser beams is significantly reduced (from 100 to 5).
We also investigate the quality of prepared state in a noisy environment, and show that the engineered code state  is indeed robust against leakage, relaxation and decoherence. 

 \begin{figure*}
\centerline{\includegraphics[width=1.0\linewidth]{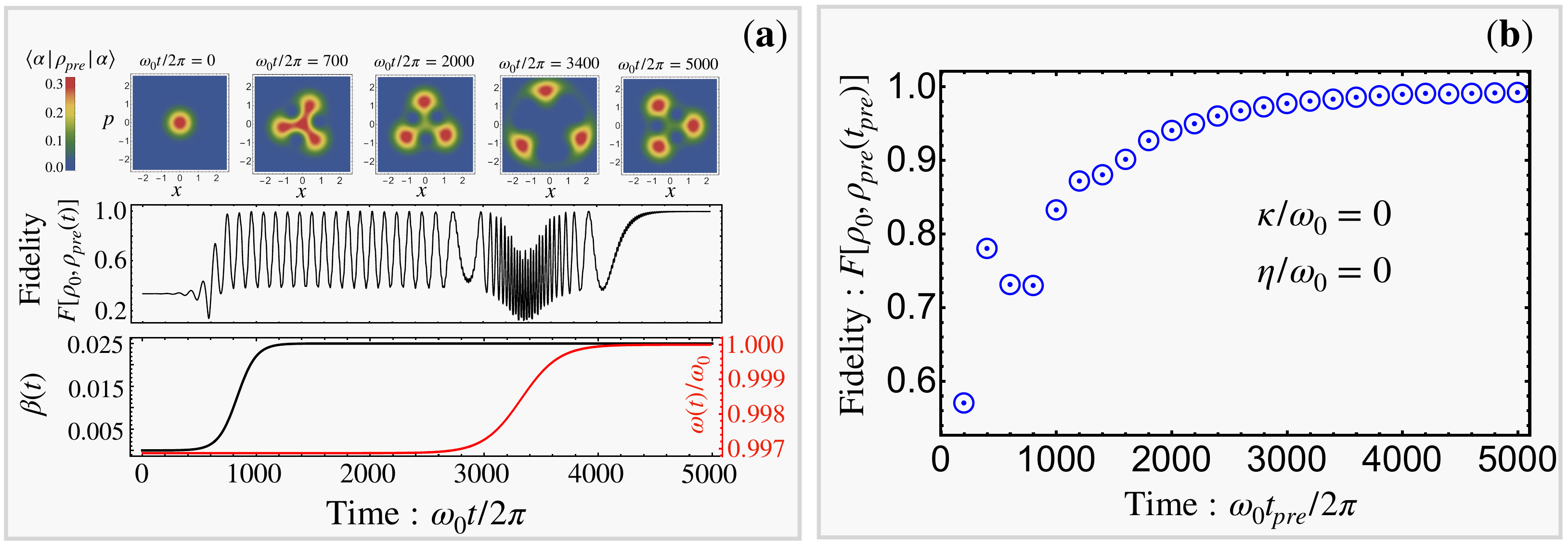}}
\caption{\label{Fig-sm-state}
{\bf State preparation with adiabatic ramp protocol.}
{\bf (a)} Lower panel: ramp for the driving amplitude $\beta(t)$ (black curve) and the driving frequency $\omega(t)$ (red curve) with the preparation time $t_{pre}=5000\times2\pi/\omega_0$; Middle panel:  stroboscopic time evolution of fidelity $F[\rho_0,\rho_{pre}(t)]$,  cf. Eq.~(\ref{eq-sm-fidelity}), of prepared state $\rho_{pre}(t)$  with respect to the 3-fold rotational target bosonic state $\rho_0=|\psi_0\rangle\langle\psi_0|$ given by Eq.~(\ref{eq-sm-code0}); Upper panel: snapshots of Husmi Q-functions of prepared states $\langle\alpha|\rho_{pre}|\alpha\rangle$ at time moments $t=0, \ 700\times2\pi/\omega_0,\  2000\times2\pi/\omega_0,\  3400\times2\pi/\omega_0$ and $5000\times2\pi/\omega_0$. {\bf (b)} Fidelity of the final prepared state $F[\rho_0,\rho_{pre}(t_{pre})]$ as a function of preparation time $t_{pre}$ without photon loss ($\kappa=0$) and pure dephasing ($\eta=0$).
 }
\end{figure*}


\subsection{~Adiabatic ramp}

To prepare the target bosonic code states, we adopt the recently proposed ``adiabatic ramp" method introduced in Ref.~\cite{xanda2023arxiv}. Following this method, we set the system Hamiltonian with periodic driving potential as
\bea\label{eq-sm-Vramp}
\hat{\mathcal{H}}(t)=
\frac{\omega_0}{2}(\hat{p}^2+\hat{x}^2)+\beta(t)V_\gamma(x,\omega(t)t),
\eea
  where $V_\gamma$ takes the form given by Eq.~(\ref{eq-sm-Vxt-1_mn})  with \emph{time-dependent} driving amplitude $\beta(t)$ and frequency $\omega(t)$.   
 The main idea is that the driving potential in Eq.~(\ref{eq-sm-Vramp}) is turned on adiabatically from $\beta(0)=0$, $\omega(0)\neq\omega_0$ within a preparation time $t_{pre}$ to the values of $\beta(t_{pre})=\beta_0$, $\omega(t_{pre})=\omega_0$. In such a way, the initial cavity state is adiabatically ramped to the target code states.   
 Following the ramp receipt present in Ref.~\cite{xanda2023arxiv},  we modulate the driving amplitude $\beta(t)$ and frequency $\omega(t)$ in form of the sigmoidal function 
 \bea
 g(t)=\frac{C}{1+e^{-s(t-t_c)}}+D
 \eea
with the center of ramp $t_c$, the slope of ramp $s$. The parameters C and D are determined by the boundary conditions 
\bea
C=\frac{g(t_{pre})-g(0)}{[1+e^{-s(t_{pre}-t_c)}]^{-1}-(1+e^{st_c})^{-1}},\ \ \ \ D=g(0)-\frac{g(t_{pre})}{1+e^{st_c}}.
\eea
As in Ref.~\cite{xanda2023arxiv}, we set $g(0)=\beta(0)=0$, $g(t_{pre})=\beta(t_{pre})=0.025\omega_0$, $t_c=t_{pre}/6$ and $s=60/t_{pre}$ for getting the profile of driving amplitude $\beta(t)$. Note that the Floquet adiabaticity  may be lost when two Floquet states have equal quasienergies modulo $\omega_0=2\pi/T$ \cite{xanda2023arxiv}. This problem can be circumvented by making the driving frequency $\omega(t)$ incommensurate with but close to the harmonic frequency \cite{xanda2023arxiv}. Here, we set $g(0)=\omega(0)=\omega_0/(1+\pi\times 10^{-3})$, $g(t_{pre})=\omega_0$, $t_c=2t_{pre}/3$ and $s=30/t_{pre}$ to fix the profile of driving frequency $\omega(t)$.   

In the lower panel of Fig.~\ref{Fig-sm-state}(a), we plot the time-varying driving amplitude $\beta(t)$ and frequency $\omega(t)$ for a preparation time $t_{pre}=5000T$. 
The preparation process starts from an initial cavity state $|\psi_{pre}(0)\rangle$, and the prepared state $|\psi_{pre}(t)\rangle$ is obtained by Schroedinger equation $$i\lambda\frac{\partial}{\partial t}|\psi_{pre}(t)\rangle=\hat{\mathcal{H}}(t)|\psi_{pre}(t)\rangle.$$  Note that the three code states given by Eqs.~(\ref{eq-sm-code0})-(\ref{eq-sm-code2}) are three different eigenstates of phase-space rotational operator $\hat{R}_q\equiv e^{i\hat{a}^\dagger\hat{a}\frac{2\pi}{q}}$ with different parity $m$, which is kept unchanged during the adiabatic preparation process by our designed driving potential. According to Eq.~(\ref{eq-sm-psilm}), the code states $|\psi_0\rangle$, $|\psi_1\rangle$ and $|\psi_2\rangle$ can be adiabatically achieved from Fock states $|0\rangle$, $|2\rangle$ and $|1\rangle$ respectively. As the cavity vacuum state is typically easier to start with than other Fock states, we focus on preparing the code state  $|\psi_0\rangle$ given by Eq.~(\ref{eq-sm-code0}) below.

To show the quality of prepared state, we define the fidelity $F[\rho_0,\rho_{pre}(t)]$  of prepared state $\rho_{pre}(t)$  with respect to the target code state $\rho_0=|\psi_0\rangle\langle\psi_0|$ by \cite{james2001pra}
\bea\label{eq-sm-fidelity}
F[\rho_0,\rho_{pre}(t)]=\mathrm{Tr}\sqrt{\rho_0^{1/2}\rho_{pre}(t)\rho_0^{1/2}}=\sqrt{\mathrm{Tr}[\rho_{pre}(t)\rho_0]}.
\eea
In the middle panel of Fig.~\ref{Fig-sm-state}(a), we plot the time evolution of fidelity calculated at the stroboscopic time moments ($\omega_0t/2\pi\in\mathbb{Z}^+$). It clearly shows that the prepared state starts with a low fidelity ($F<0.4$) and approaches the target state with high fidelity ($F>0.99$). In the upper panel of Fig.~\ref{Fig-sm-state}(a), we plot several snapshots of Husmi Q-functions of prepared states. From an initial vacuum state of cavity, the prepared state begins transiting to the target state when the driving amplitude $\beta(t)$ starts to ramp up around $t=700T$. Thereafter, the prepared state actually already achieves the target state with high fidelity but keeps oscillating due to the finite detuning ($\omega(t)\neq\omega_0$). When the driving amplitude $\beta(t)$ is close to the final value $0.025\omega_0$, the driving frequency $\omega(t)$ starts to ramp up and the fidelity of prepared state becomes lower (see the snapshot at $t=3400T$). Finally, when the driving frequency $\omega(t)$ approaches the value of $\omega_0$, the prepared state is stabilized to the target state with high fidelity again.

According to the Floquet adiabatic condition given in Ref.~\cite{xanda2023arxiv}, the quality of prepared state becomes better when the preparation time is longer. In Fig.~\ref{Fig-sm-state}(b), we plot the fidelity of final prepared state $F[\rho_0,\rho_{pre}(t_{pre})]$ as a function of preparation time $t_{pre}$. The plot verifies that the quality of prepared state is already very good when $t_{pre}>3000T$. For the typical cavity frequency $\omega_0/2\pi=5\mathrm{GHz}$ in the superconducting circuits, the preparation time is less than $1\mu s$ which is much faster than the preparation time using other protocols in circuit QED\cite{Campagne-Ibarcq2020nature,Eickbusch2022NP,Sivak2023nature}.

 \subsection{~Discretization of wavenumbers}\label{app-VIB}

 \begin{figure*}
\centerline{\includegraphics[width=0.8\linewidth]{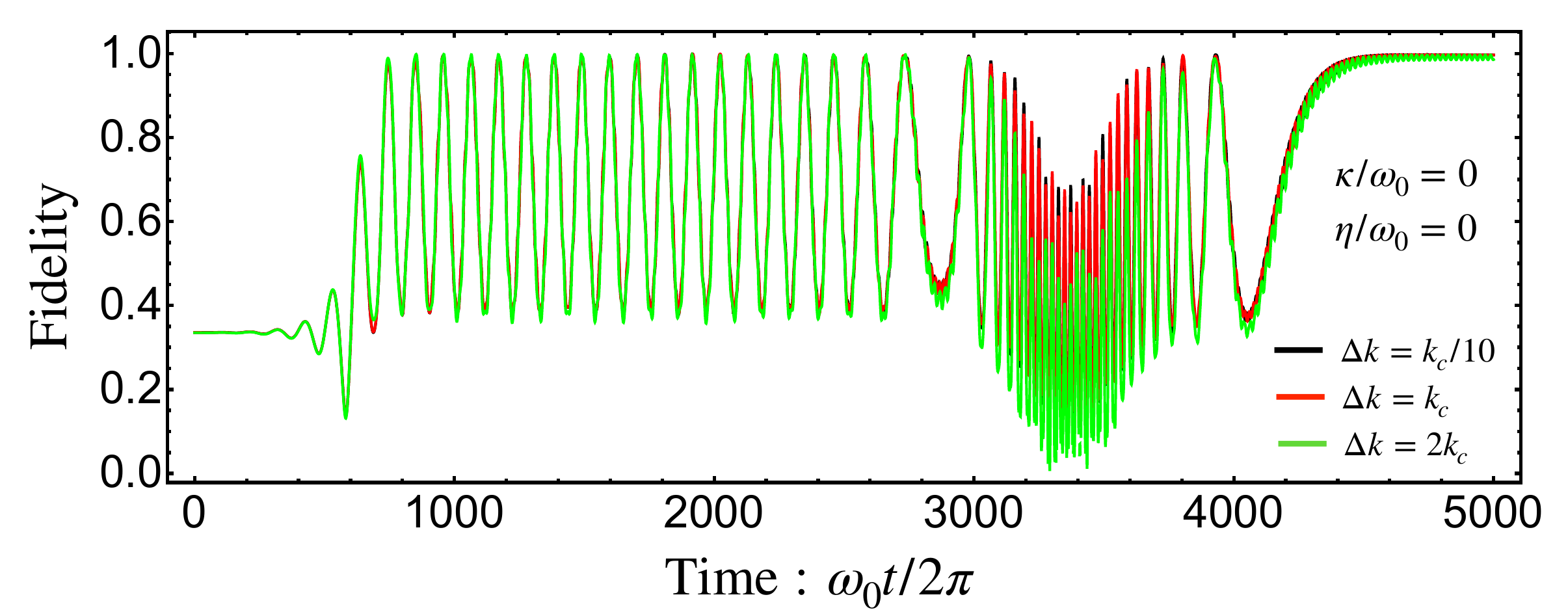}}
\caption{\label{Fig-SM-Discretization}
{\bf Dynamics with discretized wavenumbers.} Time evolution of fidelity $F[\rho_0,\rho_{pre}(t)]$ of the prepared state based on the approximate driving potential superposed by finite number of cosine functions with discretized wave numbers,  cf. Eq.~(\ref{eq-sm-Vdk}),  for different choices of discretized wavenumber step $\Delta k=k_c/10$ (black), $\Delta k=k_c$ (red) and $\Delta k=2k_c$ (Green). 
 }
\end{figure*}

We now discuss another possible errors from implementation of our driving protocol. As given by Eq.~(\ref{eq-sm-Vxt-1_mn}), the engineered driving potential is written in a  superposition of cosine potentials with modulated amplitudes and phases in time. In the real experiments, such driving potential can be created by laser beams for cold atoms \cite{Guo2022prb} or by superconducting circuits with Josephson junctions \cite{xanda2023arxiv}. In both experiments, the potential is approximated by finite number of cosine functions with discretized wavenumbers 
\bea\label{eq-sm-Vdk}
V_\gamma(x,t)&\approx&\sum_{n=1}^{k_{cut}/\Delta k}A(n\Delta k, t)\cos[(n\Delta k)x+\phi(k, t)]\Delta k.
\eea
In our numerical simulation, we choose the cutoff of wave number $k_{cut}=10k_c$, where the characteristic wavenumber is defined by $k_c\equiv\sqrt{{(1-e^{-2\gamma})}/{4\lambda}}$, cf. Eq.~(\ref{eq-sm-kc}).

In Fig.~\ref{Fig-SM-Discretization}, we plot the time evolution of fidelity for different choice of discretized wavenumber step $\Delta k=k_c/10$, $\Delta k=k_c$ and $\Delta k=2k_c$ corresponding to $100$, $10$ and $5$ tunable Josephson junctions (optical lattices) in the superconducting circuits experiment (cold atoms experiment) respectively. This means the number of Josephson junctions (laser beams) in the circuit-QED (cold atom) experiments and thus the complexity of total operations can be significantly reduced.

Our numerical results show that the discretization of the wavenumbers  causes some  discrepancies during the initial phase of the preparation and also small oscillations in the long-time behavior. Nevertheless, the averaged final fidelity of the prepared state can reach up to $99.64\%$, $99.63\%$ and $99.06\%$  for $\Delta k=0.1k_c$, $\Delta k=k_c$ and$\Delta k=2k_c$ respectively. 
The fidelity of the state prepared by using our scheme is higher than the fidelity obtained for cat states of similar amplitude using a sequence of interleaved Selective Number-dependent Arbitrary Phase (SNAP)  gates and displacement gates  \cite{Kudra2022prxq}.  
We note in passing that the infidelity obtained using our method could be further increased without adding any additional experimental complexity. In fact,
we expect the residual infidelity in our numerical results to be mainly due to   high-order Floquet-Mafnus corrections (beyond the RWA) creating a deviation between  the implemented and the target Floquet Hamiltonian. This deviation could be reduced  by fine-tuning  the driving potential to account for higher-order terms. Such an extension of our work is in preparation. 

Coherently controlling multiple tunable Josephson Junctions (JJs) for designing functional quantum devices and quantum computation/simulation is a well-established technology in circuit-QED architectures. Examples include the Josephson ring modulator architecture \cite{Bergeal2010np,roch2012prl} with $4$ JJ (one for each transmon qubit), the quantum-state-preservation superconducting circuit \cite{Kelly2015nature} with $9$ transmons, the Google programmable superconducting processor Sycamore \cite{Arute2019nat} with $54$ transmon qubits  and the recent IBM quantum processor Eagle \cite{Kim2023nat} with $127$ transmons qubit. We also note in passing that, in the spirit of Trotter discretization \cite{seth1996sci}, our driving scheme could even be realized even with a single transmon by decomposing the multiple JJs unitary operation into a sequence of discrete gate operations. We leave a detailed study of this scenario to a future work \cite{guo2024}.

\subsection{~State preparation in noisy environment}\label{sec-sm-noise}

As discussed in the main text, our proposal can be realized with ultracold atom in optical lattices \cite{Moritz2003prl,Hadzibabic2004prl,Guo2022prb}  or superconducting circuits with Josephson junctions \cite{xanda2023arxiv,Chen2014prb,Hofheinz2011prl,Chen2011apl,Armour2013prl,Armour2015prb,Trif2015prb,Hofer2016prb,Dambach2017njp,Lang2022arxiv}. Compared to clean ultracold atomic systems, superconducting circuits are interacting with dirtier environment because of charge and flux noises. For typical superconducting aluminium cavity with frequency $\omega_0/2\pi=5\mathrm{GHz}$, the cavity relaxation time $T_1=1/\kappa$ can be longer than $500\mu s$ \cite{Eickbusch2022NP,Sivak2023nature} corresponding to $\kappa/\omega_0\approx 6.4\times 10^{-8}$. Usually, the pure dephasing rate of cavity is much weaker than the relaxation rate with typical pure dephasing time $T_{\phi}=1/\eta$ longer than $1ms$ \cite{Eickbusch2022NP,Sivak2023nature} meaning $\eta/\omega_0 < 3.2\times 10^{-8}$. The dynamics of prepared state in a noisy environment is described by the Lindblad master equation (\ref{eq-sm-rho}).

\begin{figure*}
\centerline{\includegraphics[width=0.9\linewidth]{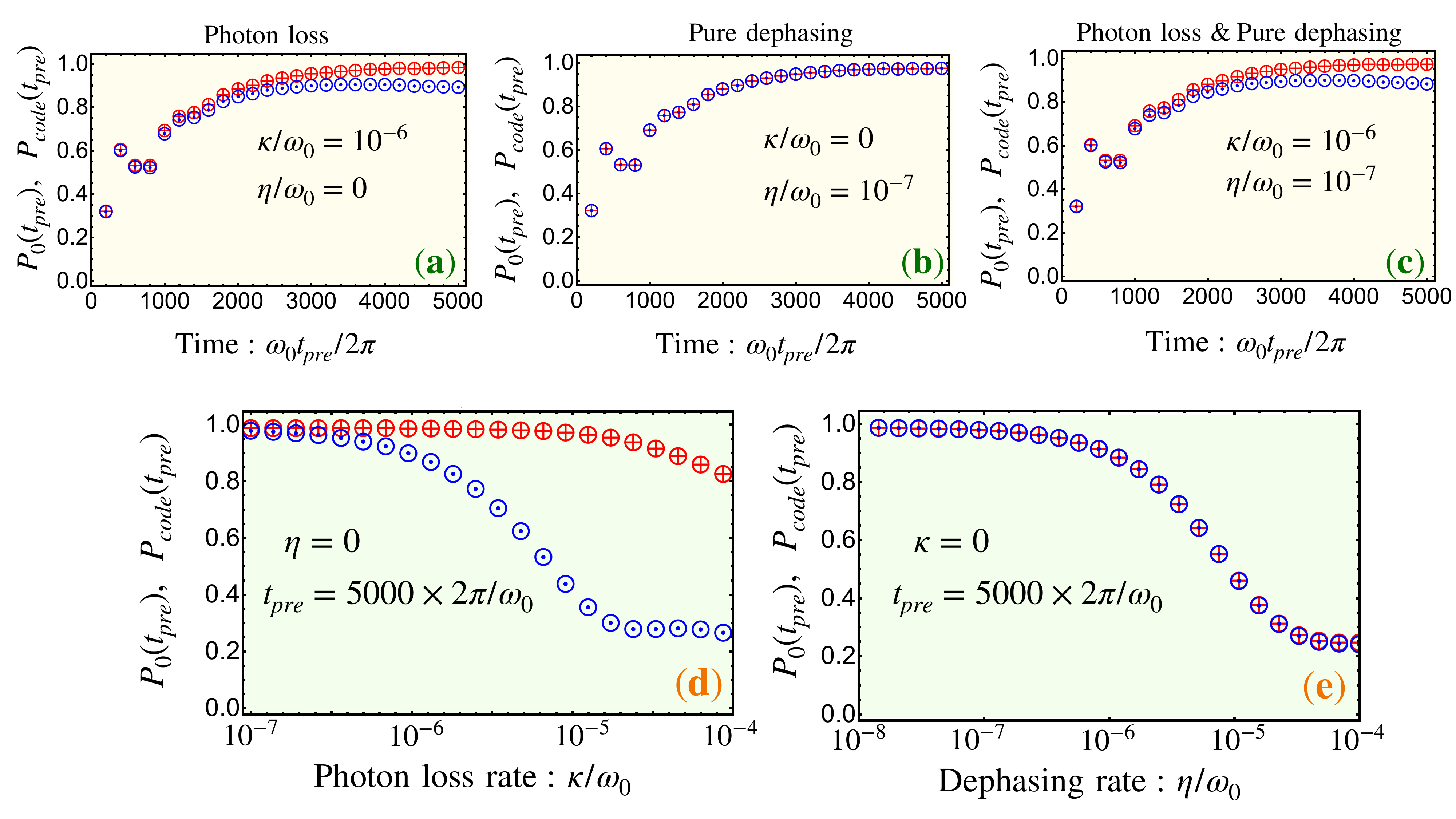}}
\caption{\label{Fig-sm-state-noisy}
{\bf Quality of prepared states in noisy environment.}
{\bf (a)-(c)} Probability of the prepared state over the target state $P_0$ (blue circled dots ), cf. Eq.~(\ref{eq-sm-P0}), and total probability in the code space $P_{code}$ (red circled pluses), cf. Eq.~(\ref{eq-sm-Pcode}), as functions of preparation time $t_{pre}$ for different photon loss rates $\kappa/\omega_0$ and pure dephasing rates $\eta/\omega_0$. {\bf (d)-(e)}  $P_0(t_{pre})$ and $P_{code}(t_{pre})$ of the prepared state at final time $t_{pre}=5000\times 2\pi/\omega_0$ as functions of photon loss rate $\kappa/\omega_0$ and pure dephasing rate $\eta/\omega_0$.
 }
\end{figure*}

\subsection{~Photon loss}

We first set pure dephasing rate $\eta=0$ and discuss the effects of finite photon loss rate $\kappa>0$. To analyse the errors from photon loss during state preparation, we unravel the master equation Eq.~(\ref{eq-sm-rho}) with wave-function Monte Carlo method \cite{jacobs2010pra} in the framework of quantum trajectory theory \cite{Wiseman2009book}. For an initial pure quantum state, each stochastic quantum trajectory of prepared state $|\psi^{(n)}_{pre}(t)\rangle$ can be constructed by the following map for any infinitesimal time interval  $dt$ \cite{jacobs2010pra}
\bea\label{eq-sm-qt}
 |\psi^{(n)}_{pre}(t+dt)\rangle=\Big[1-\frac{i}{\lambda}\hat{\mathcal{H}}(t)dt\Big]|\psi^{(n)}_{pre}(t)\rangle-\frac{1}{2}\kappa dt\big(\hat{a}^\dagger\hat{a}-2\langle \hat{a}^\dagger+\hat{a}\rangle_n\hat{a}\big)|\psi^{(n)}_{pre}(t)\rangle+\sqrt{\kappa}dW_n\hat{a}|\psi^{(n)}_{pre}(t)\rangle
\eea
together with a normalization procedure of $ |\psi^{(n)}_{pre}(t+dt)\rangle$. Here, $\langle \hat{a}^\dagger+\hat{a}\rangle_n=\langle \psi^{(n)}_{pre}(t)|(\hat{a}^\dagger+\hat{a})|\psi^{(n)}_{pre}(t)\rangle$
and $dW_n$ is Wiener noise increment satisfying $(dW_n)^2=dt$ independent of each quantum trajectory. The density matrix of the prepared state at time $t$ is approximated by $N$ quantum trajectories $$\rho_{pre}(t)\approx\frac{1}{N}\sum_{n=1}^N |\psi^{(n)}_{pre}(t)\rangle \langle\psi^{(n)}_{pre}(t)|.$$  
The first term on the right hand side (RHS) of Eq.~(\ref{eq-sm-qt}) represents the unitary time evolution by adiabatic ramp. The second and third terms on the RHS of Eq.~(\ref{eq-sm-qt}) describe dissipation and fluctuation (quantum jump) respectively. 

In the weak photon loss rate regime ($\sqrt{\kappa}dW\gg\kappa dt$), the jump term is dominant over the dissipation term. For the rotational code state $|\psi_{l,m}\rangle$ given by Eq.~(\ref{eq-sm-psilm}),  the consequence from quantum jump is 
 \bea\label{eq-sm-jump}
 \hat{a}|\psi_{l,m}\rangle\propto\sum_{p=0}^{q-1}e^{imp\frac{2\pi}{q} }\hat{a}\hat{R}_q^p|\phi_l\rangle
 \propto\sum_{p=0}^{q-1}e^{i(m+1)p\frac{2\pi}{q} }\hat{R}_q^p\hat{a}|\phi_l\rangle
 \propto|\psi_{l,m+1}\rangle.
 \eea 
Here, we have used the property $\hat{R}_q^{\dagger }\hat{a}\hat{R}_q=\hat{a}e^{i\frac{2\pi}{q}}$ in the second step, and $\hat{a}|\phi_l\rangle=\alpha_0e^{-\gamma}|\phi_l\rangle$ with $\phi_l=\alpha_0e^{-\gamma}$ in the last step. The relation $\hat{a}|\psi_{l,m}\rangle\propto|\psi_{l,m+1}\rangle$ means that the single-photon loss only alternates the Bloch eigenstates inside the ``same" band without inter-band transition, and thus does not leak quantum information out of the code subspaces \cite{Gottesman2001pra,Puri2019PRX,Rymarz2021prx,Conrad2021pra}. 

The quantum trajectory Eq.~(\ref{eq-sm-qt}) is used to analyze the effects of noises during preparation. To verify the above analysis, we obtain numerical results directly from master equation (\ref{eq-sm-rho}). We define the probability of the prepared state over the target state 
\bea\label{eq-sm-P0}
P_0(t)\equiv \langle\psi_0|\rho_{pre}(t)|\psi_0\rangle.
\eea
and compare it to the total probability in the code space $P_{code}(t)$ defined by Eq.~(\ref{eq-sm-Pcode}).
In Fig.~\ref{Fig-sm-state-noisy}(a), we plot  $P_0(t_{pre})$ and $P_{code}(t_{pre})$ of prepared state as function of preparation time $t_{pre}$ with finite photon loss rate $\kappa/\omega_0=10^{-6}$. As expected, although the probability $P_0(t_{pre})$ of prepared state over the target state at final time $t_{pre}=5000\times 2\pi/\omega_0$ is indeed lowered a bit, 
the  total probability in the code space $P_{code}(t_{pre})$ for long reparation time $t_{pre}=5000\times 2\pi/\omega_0$ is close to one.  

In Fig.~\ref{Fig-sm-state-noisy}(d), we plot $P_0(t_{pre})$ and $P_{code}(t_{pre})$ of prepared state at final time $t_{pre}=5000\times 2\pi/\omega_0$ as function of photon loss rate $\kappa/\omega_0$. 
Our result shows that  the quantum information is well protected inside the code space, i.e., $P_{code}(t_{pre})$ keeps close to one, as long as $\kappa/\omega_0<10^{-5}$ which is satidfied in the real circuit-QED experiments \cite{Eickbusch2022NP,Sivak2023nature}.
For strong photon loss rate $\kappa/\omega_0>10^{-5}$, the quantum information will start to leak outside the code space due to the nonnegligible excitation of prepared state into high-band states ($P_{code}<1$). The main channel is $\hat{a}^\dagger \hat{a}|\psi_{l,m}\rangle$ in the dissipation term on the RHS of Eq.~(\ref{eq-sm-qt}), i.e.,
\bea\label{eq-sm-adga}
\hat{a}^\dagger \hat{a}|\psi_{l,m}\rangle\propto\sum_{p=0}^{q-1}e^{imp\frac{2\pi}{q} }\hat{a}^\dagger\hat{a}\hat{R}_q^p|\phi_l\rangle
 \propto\sum_{p=0}^{q-1}e^{imp\frac{2\pi}{q} }\hat{R}_q^p\hat{a}^\dagger\hat{a}|\phi_l\rangle. 
\eea
In the second step, we have used the property $\hat{R}_q^{\dagger }\hat{a}^\dagger\hat{a}\hat{R}_q=\hat{a}^\dagger\hat{a}$. Because $\hat{a}^\dagger\hat{a}|\alpha\rangle=\alpha\hat{a}^\dagger|\alpha\rangle$ is no longer a standard coherent state,  the resultant state $\hat{a}^\dagger \hat{a}|\psi_{l,m}\rangle$ can be a superposition of the states with the same parity $m$ in all Bloch bands. Thus, the operator $\hat{a}^\dagger \hat{a}$ can introduce inter-band transition making the quantum information leaking outside of the code space. 
Other channels introducing such leakage errors may include the non-RWA effects, the adiabatic approximation and the deformation of the cell state $|\phi_l\rangle$ from the standard coherent state due to the finite detuning $(\omega(t)\neq\omega_0)$, cf.  the snapshot in the upper panel of Fig.~\ref{Fig-sm-state}(a) at earlier time of preparation process $t=700\times 2\pi/\omega_0$.

The photon loss leads the system state to the code subspaces without distinguishing the states in the code space. To prepare a specific target state in code space, one can detect the photon number loss and track the change of parity during preparation. As single-photon loss only alternates the Bloch index $m\rightarrow m+1$ during the whole preparation time according to Eq.~(\ref{eq-sm-jump}), we can correct such flipping errors inside the code space by \emph{updating the knowledge of quantum state when one single photon is detected}  without  backaction on the encoded system. This error correction scheme can be extended to mult-photon loss because of $\hat{a}^n|\psi_{l,m}\rangle\propto|\psi_{l,m+n}\rangle$. All in all, we only need to counter how many photons are detected in total,  and update our knowledge accordingly in the end end of preparation process.  As long as the efficiency of photon number detection is perfect, the errors from photon loss can be $100\%$ tracked and corrected. Alternatively, one can directly perform parity measurement and post select the code state \cite{Terhal2020iop}.

 \subsection{~Pure dephasing}
 
 Now, we set photon loss rate $\kappa=0$ and discuss the effects of finite pure dephasing rate $\eta>0$. In this case, the stochastic quantum trajectory of prepared state $|\psi^{(n)}_{pre}(t)\rangle$ can be constructed by the following map \cite{jacobs2010pra}
\bea\label{eq-sm-qteta}
 |\psi^{(n)}_{pre}(t+dt)\rangle=\Big[1-\frac{i}{\lambda}\hat{\mathcal{H}}(t)dt\Big]|\psi^{(n)}_{pre}(t)\rangle-\frac{1}{2}\eta dt\big[(\hat{a}^\dagger\hat{a})^2-4\langle \hat{a}^\dagger\hat{a}\rangle_n\hat{a}^\dagger\hat{a}\big]|\psi^{(n)}_{pre}(t)\rangle+\sqrt{\eta}dW_n\hat{a}^\dagger\hat{a}|\psi^{(n)}_{pre}(t)\rangle
\eea
together with a normalization procedure of $ |\psi^{(n)}_{pre}(t+dt)\rangle$, where $\langle \hat{a}^\dagger\hat{a}\rangle_n=\langle \psi^{(n)}_{pre}(t)|(\hat{a}^\dagger\hat{a})|\psi^{(n)}_{pre}(t)\rangle$.
 Accordding to Eq.~(\ref{eq-sm-adga}), starting from a rotational state $|\psi_{l,m}\rangle$, the noisy terms $\hat{a}^\dagger \hat{a}|\psi_{l,m}\rangle$ and $(\hat{a}^\dagger \hat{a})^2|\psi_{l,m}\rangle$ on the RHS of Eq.~(\ref{eq-sm-qteta}) keep the parity $m$ of prepared state exactly unchanged during the whole preparation process. As a result, the probability of the prepared state over the target state $P_0$ and the total probability in the code space $P_{code}$ must be exactly the same  (no state flipping in the code space). 
 
 In Fig.~\ref{Fig-sm-state-noisy}(b), we plot  $P_0(t_{pre})$ and $P_{code}(t_{pre})$ of prepared state as a function of preparation time $t_{pre}$ with finite pure dephasing rate $\eta/\omega_0=10^{-7}$ and zero photon loss $\kappa/\omega_0=0$. It clearly shows that the probability of prepared state over the target state $P_0(t_{pre})$ coincides with total probability in the code space $P_{code}(t_{pre})$ for any preparation time. 
 In Fig.~\ref{Fig-sm-state-noisy}(e), we further plot  $P_0(t_{pre})$ and $P_{code}(t_{pre})$ of prepared state at final time $t_{pre}=5000\times 2\pi/\omega_0$ as function of pure dephasing rate $\eta/\omega_0\in (10^{-8},10^{-4})$. The probabilities $P_0(t_{pre})$ and $P_{code}(t_{pre})$ still coincide with each other showing that the pure dephasing noise does not introduce intra-band transition. 
 
 Our code preparation protocol is robust as long as $\eta/\omega_0<10^{-7}$ corresponding to pure dephasing time $T_{\phi}<0.32 ms$ for a microwave superconducting cavity with frequency $\omega_0/2\pi=5\mathrm{GHz}$. The significant leakage of quantum information outside the code space for  $\eta/\omega_0>10^{-7}$ comes from the consequence of $\hat{a}^\dagger \hat{a}|\psi_{l,m}\rangle$ and $(\hat{a}^\dagger \hat{a})^2|\psi_{l,m}\rangle$ in the dissipation term on the RHS of Eq.~(\ref{eq-sm-qteta}) as discussed below Eq.~(\ref{eq-sm-adga}).  %
Finally,  in Fig.~\ref{Fig-sm-state-noisy}(c), we plot  $P_0(t_{pre})$ and $P_{code}(t_{pre})$ of prepared state as a function of preparation time $t_{pre}$ with finite photon loss rate $\kappa/\omega_0=10^{-6}$ and finite pure dephasing rate $\eta/\omega_0=10^{-7}$ showing that our code preparation protocol is indeed robust against decoherence in the noisy environment.

\subsection{~Preparation of four-fold code states }

\begin{figure*}
\centerline{\includegraphics[width=\linewidth]{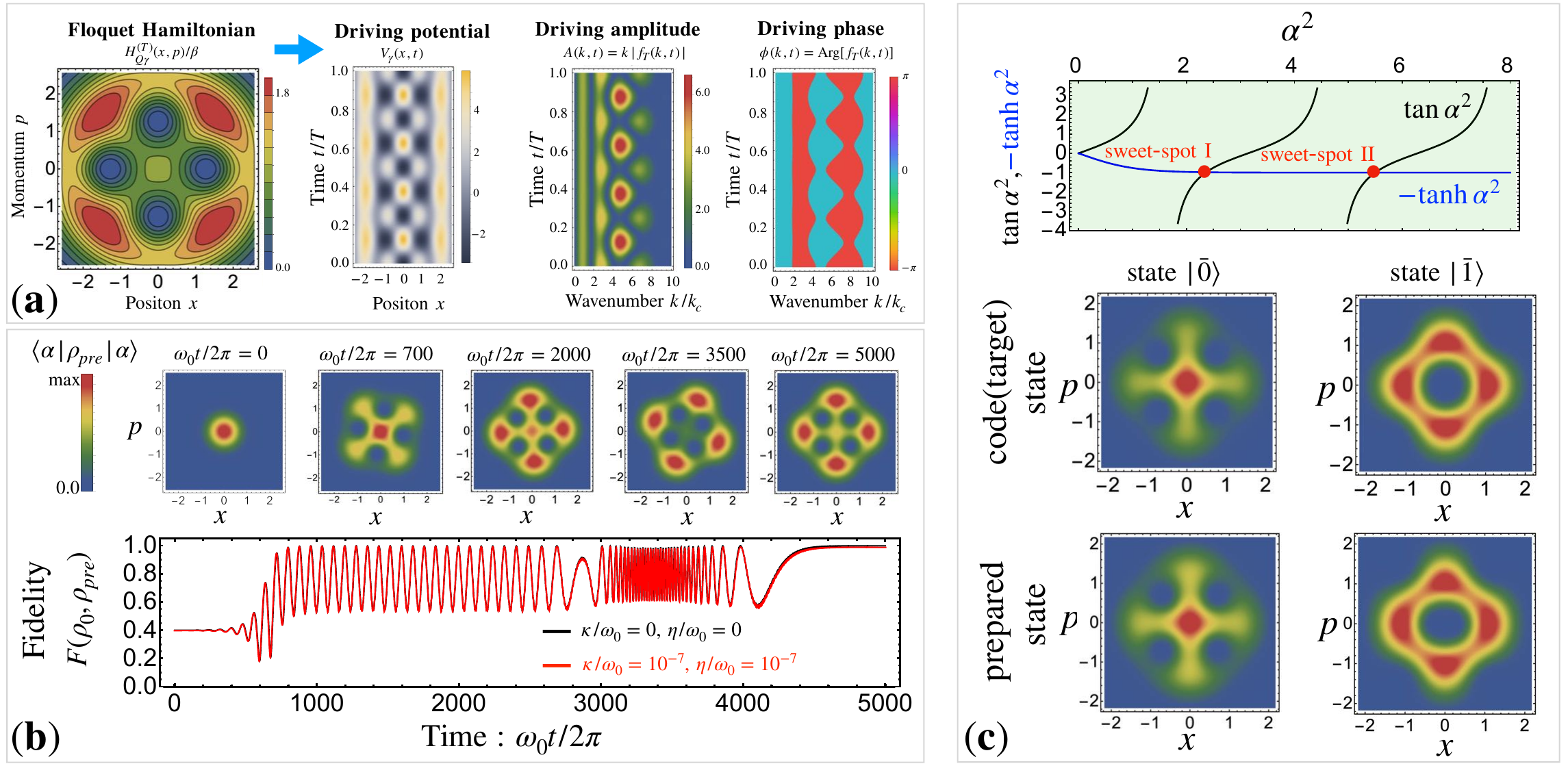}}
\caption{\label{Fig-SM-State-4-6}
{\bf Preparation of four-fold code states.}  {\bf (a)} Engineering $4$-fold rotational symmetric Hamiltonian given by Eq.~(\ref{eq-sm-HTq}) with parameters: $q=4$, $\lambda=1/4$, $\alpha_0=\sqrt{2}$ and $\gamma=1/4$.  Subfigures from left to right: Hamiltonian Q-function $H^{(T)}_{Q\gamma}(x,p)/\beta$, real-space driving potential $V_\gamma(x,t)$, time-varying amplitude $A(k,t)$ and phase $\phi(k,t)$ of cosine component of engineered driving potential, cf. Eq.~(\ref{eq-sm-Vxt-1_mn}).  
{\bf (b)} Lower panel: Time evolution of fidelity $F[\rho_0,\rho_{pre}(t)]$ of prepared state $\rho_{pre}(t)$  with respect to the 4-fold rotational target bosonic state $\rho_0=|\psi_0\rangle\langle\psi_0|$ given by Eq.~(\ref{eq-sm-psimcode}) in clean (black curve, $\kappa=0,\ \eta=0$) and noisy (red curve, $\kappa/\omega_0=10^{-7},\ \eta/\omega_0=10^{-7}$) environment; Upper panel: snapshots of Husmi Q-functions $\langle\alpha|\rho_{pre}(t)|\alpha\rangle$  of prepared states at time moments $t=0, \ 700\times 2\pi/\omega_0,\  2000\times 2\pi/\omega_0,\  3500\times 2\pi/\omega_0$ and $5000\times 2\pi/\omega_0$ in clean environment. 
{\bf (c)}  Preparation of four-legs cat code words $|\bar{0}\rangle$ and $|\bar{1}\rangle$. The upper panel show the functions $\tan\alpha^2$ and $-\tanh\alpha^2$ as functions of $\alpha^2$ with \emph{sweet spots} (red dots) satisfying quantum error correction condition (\ref{eq-sm-kl}). We choose the parameters $q=4$, $\lambda=1/4$, $\gamma=1/4$, and set $\alpha=\alpha_0e^{\gamma}$ at sweep spot I for plotting the Husmi-Q functions of code-word states (lower panel). The adiabatic ramp parameters are the same as that for figure (b) and the prepared code states are obtained at final moment $t_{pre}=5000\times 2\pi/\omega_0$.
 }
\end{figure*}

Below, we introduce the engineering of the four-fold symmetric cat code states that were introduced in Refs.~\cite{Mirrahimi2014njp,Leghtas2013prl}. The codewords can be obtained from Eq.~(\ref{eq-sm-psimcode})  with the following four-legs form \cite{Terhal2020iop}
\bea
&&|\bar{0}\rangle\equiv|\psi_{m=0}\rangle=\frac{1}{\sqrt{\mathcal{N}_0}}\Big(|\alpha\rangle+|-\alpha\rangle+|i\alpha\rangle+|-i\alpha\rangle\Big)\label{eq-sm-4cat1}\\
&&|\bar{1}\rangle\equiv|\psi_{m=2}\rangle=\frac{1}{\sqrt{\mathcal{N}_2}}\Big(|\alpha\rangle+|-\alpha\rangle-|i\alpha\rangle-|-i\alpha\rangle\Big)\label{eq-sm-4cat2},  
\eea
where $\alpha=\alpha_0e^\gamma\in\mathbb{R}$ and $\mathcal{N}_{m}=8e^{-\alpha^2}(\cosh\alpha^2+(-1)^{\frac{m}{2}}\cos\alpha^2)$. 
Such four-legs cat code states can be prepared by our phase-space Hamiltonian engineering method combined with adiabatic ramp. 

In Fig.~\ref{Fig-SM-State-4-6}(a), we first engineer a $4$-fold rotational symmetric Hamiltonian $\hat{H}^{(T)}_{F\gamma}$ that is given by Eq.~(\ref{eq-sm-HTq}) with $q=4$ whose zero-energy eigenstates are the above code-word states.  The subfigures from left to right represent the Hamiltonian Q-function $H^{(T)}_{Q\gamma}(x,p)/\beta$, the real-space driving potential $V_\gamma(x,t)$, the time-varying amplitude $A(k,t)$ and the phase $\phi(k,t)$ of cosine component of designed driving potential, cf. Eq.~(\ref{eq-sm-Vxt-1_mn}).  The code state we aim to prepare is the code word $|\bar{0}\rangle$ given by Eq.~(\ref{eq-sm-4cat1}). In the lower panel of Fig.~\ref{Fig-SM-State-4-6}(b), we show the time evolution of fidelity $F[\rho_0,\rho_{pre}(t)]$ of prepared state $\rho_{pre}(t)$ with respect to $\rho_0=|\bar{0}\rangle\langle\bar{0}|$ from cavity vacuum state with the adiabatic ramp method (lower panel) in the clean (black curve) and noisy (red curve) environment. In the upper panel of Fig.~\ref{Fig-SM-State-4-6}(b), we show the snapshots of Husmi Q-functions of prepared states $\langle\alpha|\rho_{pre}(t)|\alpha\rangle$ at different time moments in the clean environment. Our results show that the preparation process is indeed robust to the noises in the environment.

One can verify that the above code-word states given by Eqs.~(\ref{eq-sm-4cat1}) and (\ref{eq-sm-4cat2}) are the  eigenstates of the photon parity operator $\hat{\prod}\equiv e^{i\pi\hat{a}^\dagger\hat{a}}$ with eigenvalues ``+1". The code states $|\bar{0}\rangle$  and $|\bar{1}\rangle$, however, are distinguished by the operator $\hat{\prod}^{\frac{1}{2}}\equiv e^{i\frac{\pi}{2}\hat{a}^\dagger\hat{a}}$ since $\hat{\prod}^{\frac{1}{2}}|\bar{0}\rangle=|\bar{0}\rangle$ and $\hat{\prod}^{\frac{1}{2}}|\bar{1}\rangle=-|\bar{1}\rangle$.  It is natural to detect photon loss and extend the lifetime of the above bosonic code states by  tracking the photon parity from repeated measurements of $\hat{\prod}$ operator. Such error correction scheme has been realized using  a transmon qubit and cavity mode \cite{Ofek2016nature,Schoelkopf2018science}. However, to ideally correct the single-photon loss error, the chosen four-legs cat state need satisfy the Knill-Laflamme condition \cite{chuang2010book}: $\langle \bar{0}|\hat{a}^\dagger\hat{a}|\bar{0}\rangle=\langle \bar{1}|\hat{a}^\dagger\hat{a}|\bar{1}\rangle$, which turns to be \cite{Terhal2020iop}
\bea\label{eq-sm-kl}
\tan\alpha^2=-\tanh\alpha^2
\eea
In the upper panel of Fig.~\ref{Fig-SM-State-4-6}(c), we plot the functions $\tan\alpha^2$ and $-\tanh\alpha^2$ as functions of $\alpha^2$. The crossing points that meet the above quantum error condition are called \emph{sweet spots}. The four colored plots below show the Husmi-Q functions of code-word states at sweet spot I, where the first row shows the target code states while the second row shows the prepared code states. The code state $|\bar{0}\rangle$ is prepared from the vacuum cavity state while the other code state $|\bar{1}\rangle$ is prepared from the second excited Fock state of undriven cavity.


\section{~Sharp-boundary elliptical well in phase space}\label{app-VII}

In this section, we 
calculate analytical expression of the engineered driving potential for the elliptical Hamiltonian in phase space. We start from the general formula of the engineered driving potential as follows
\bea\label{Vxt}
V(x,t)=\frac{1}{2}\int_{-\infty}^{+\infty} |k|f_T(k,t)e^{i kx}dk.
\eea
According to the identity
$
x^{-2}=-\frac{1}{2}\int_{-\infty}^{+\infty} |k|e^{ikx}dk
$ and defining the following function
\bea\label{pxt}
v(x,t)\equiv\frac{1}{2\pi}\int_{-\infty}^{+\infty}f_T(k,t)e^{ikx}dk,
\eea
the driving potential $V(x,t)$ is the convolution of $-x^{-2}$ and $u(x,t)$, i.e.,
\bea\label{eq-SM-Vxt}
V(x,t)&=&-x^{-2}*v(x,t)\nl
&=&-\int_{-\infty}^{+\infty}\frac{1}{z^2}v(x-z,t)dz\nl
&=&-\lim_{\epsilon \to 0^+} \int_{-\infty}^{+\infty}\Re\Big[\frac{1}{(z-i\epsilon)^2}\Big]v(x-z,t)dz\nl
&=&-\lim_{\epsilon \to 0^+} \int_{-\infty}^{+\infty}\frac{z^2-\epsilon^2}{(z^2+\epsilon^2)^2}v(x-z,t)dz.
\eea
Here, we should first replace the convolution function $1/z^2$ by ${(z^2-\epsilon^2)}/{(z^2+\epsilon^2)^2}$ to get converged integral, and then take the limit of $\epsilon \to 0$ to obtain $V(x,t)$.

In the $(x,p)$ phase space, we set a new coordinate $(x',p')$ system which is rotated by an angle $\tau$ between $x$ and $x'$ axes given by the following orthogonal transformation
\bea\label{eq-sm-xx'pp'}
x=x'\cos\tau-p'\sin\tau; \ \ \ p=x'\sin\tau+p'\cos\tau.
\eea
We can express the target Hamiltonian Q-function with rotated coordinates by $H^{'(T)}_Q(x',p')\equiv H^{(T)}_Q(x,p)$. Then, we project the Hamiltonian on the $x'$ axis by the so-called \emph{Radon transformation}
\bea\label{eq-RH'}
\mathcal{R}_\tau[H^{'(T)}_Q](x')=\int _{-\infty}^{+\infty}H^{'(T)}_Q(x',p')dp'.
\eea
The 1D Fourier transformation of the above projected function is given by
\bea\label{eq-FH'}
\mathcal{F}[\mathcal{R}_\tau[H^{'(T)}_Q]](k_{x'})=\int \mathcal{R}_\tau[H^{'(T)}_Q](x')e^{-ik_{x'}x'}dx'.
\eea
The 2D Fourier transformation of $H^{'(T)}_Q(x',p')$ is 
\bea\label{eq-FH'kx'}
\mathcal{F}[H^{'(T)}_Q](k_{x'},k_{p'})=\int\int H^{'(T)}_Q(x',p')e^{-i(k_{x'}x'+k_{p'}p')}dx'dp'.
\eea
By plugging Eq.~(\ref{eq-RH'}) into Eq.~(\ref{eq-FH'}) and comparing with Eq.~(\ref{eq-FH'kx'}), we have 
\bea
\mathcal{F}[\mathcal{R}_\tau[H^{'(T)}_Q]](k_{x'})=\mathcal{F}[H^{'(T)}_Q](k_{x'},k_{p'}=0).
\eea
This is the so-called projection-slice theorem \cite{Bracewell1956ajp,Ren2005acm}. Comparing to Eq.~(\ref{eq-fFT}) and using the orthogonal transformation (\ref{eq-sm-xx'pp'}),
 we have $\mathcal{F}[\mathcal{R}_\tau[H^{'(T)}_Q]](k)=2\pi f_T(k,\tau)e^{-\frac{\lambda}{4}k^2}$ (here we have set $k_{x'}=k$) and thus
\bea\label{}
\mathcal{R}_\tau[H^{'(T)}_Q](x)=\int^{+\infty}_{-\infty} f_T(k,\tau)e^{-\frac{\lambda}{4}k^2}e^{ikx}dk.\
\eea
In the classical limit $\lambda=0$, according to Eqs.~(\ref{pxt}) and (\ref{eq-SM-Vxt}), we have $v(x,\tau)=\frac{1}{2\pi}\mathcal{R}_\tau[H^{'(T)}_Q](x)$ and thus
\bea\label{eq-sm-Vxt}
V(x,t)&=&-\frac{1}{2\pi }\frac{1}{x^2}*\mathcal{R}_{\tau=t}[H^{'(T)}_Q](x).
\eea

Now, we apply the convolutional form (\ref{eq-SM-Vxt}) and Radon transformation (\ref{eq-RH'}) to calculate the engineered driving potential for the elliptical well in phase space in the classical limit $\lambda=0$.  The boundary of the elliptical Hamiltonian Q-function in phase space is given by
$$
\frac{(x'\cos\tau-p'\sin\tau)^2}{a^2}+\frac{(x'\sin\tau+p'\cos\tau)^2}{b^2}=1.
$$
Using the transformation (\ref{eq-sm-xx'pp'}), we have the following 
$$
\Big(\frac{\sin \tau^2}{a^2}+\frac{\cos\tau^2}{b^2}\Big)p'^2-\sin (2\tau)\Big(\frac{1}{a^2}-\frac{1}{b^2}\Big)x'p'+\Big(\frac{\cos \tau^2}{a^2}+\frac{\sin\tau^2}{b^2}\Big)x'^2-1=0.
$$
Using the two solutions $p'_1$ and $p'_2$ of the above equation, the length across the ellipse is
\bea
|p'_1-p'_2|&=&\sqrt{(p'_1+p'_2)^2-4p'_1p'_2}\nl
&=&\sqrt{\frac{1}{A^2}x'^2\sin^2 2\tau \Big(\frac{1}{a^2}-\frac{1}{b^2}\Big)^2-\frac{4}{A}\Big[x'^2\Big(\frac{\cos \tau^2}{a^2}+\frac{\sin \tau^2}{b^2}\Big)-1\Big]}\nl
&=&\frac{2}{\sqrt{A}}\sqrt{1-Bx'^2}
\eea
where 
\bea
A(\tau)\equiv\frac{\sin \tau^2}{a^2}+\frac{\cos\tau^2}{b^2},\ \ \  B(\tau)\equiv \Big(\frac{\cos\tau^2}{a^2}+\frac{\sin \tau^2}{b^2}\Big)-\frac{1}{4A(\tau)}\sin^2 2\tau \Big(\frac{1}{a^2}-\frac{1}{b^2}\Big)^2
\eea
From Eq.~(\ref{eq-RH'}), we have the Radon transformation
\bea
\mathcal{R}_{\tau}[H^{'(T)}_Q](x')=-|p'_1-p'_2|=-\frac{2}{\sqrt{A(\tau)}}\sqrt{1-B(\tau)x'^2}\ \ \ \mathrm{for}\ \ \ |x'|<\frac{1}{\sqrt{B(\tau)}}.\ \ \ \ \ \ \ \ 
\eea
From Eq.~(\ref{eq-sm-Vxt}), the driving potential is given by
\bea\label{vxtellipse}
V(x,t)&=&-\frac{1}{2\pi}x^{-2}*\mathcal{R}_{t}[H^{'(T)}_Q](x)\nl
&=&-\frac{1}{2\pi}\lim_{\epsilon \to 0^+} \int_{-\infty}^{+\infty}\frac{z^2-\epsilon^2}{(z^2+\epsilon^2)^2}\mathcal{R}_{t}[H^{'(T)}_Q](x-z)dz\nl
&=&\frac{1}{2\pi}2\sqrt{\frac{B}{A}}\lim_{\epsilon \to 0^+} \int_{-\infty}^{+\infty}\frac{z^2-\epsilon^2}{(z^2+\epsilon^2)^2}\sqrt{\big(\frac{1}{\sqrt{B}}-x+z\big)\big(\frac{1}{\sqrt{B}}+x-z\big)}dz\nl
&=&-\sqrt{\frac{B}{A}}\lim_{\epsilon \to 0^+}\Big(1-\Re\big[\frac{\epsilon-ix}{\sqrt{B^{-1}+(\epsilon-ix)^2}}\big]\Big)
\eea
In the limit $\epsilon=0$, we have
\bea\label{eq-ellipVxt-sm}
V(x,t)=
\begin{cases}
   -\sqrt{\frac{B(t)}{A(t)}},&  |x|<\frac{1}{\sqrt{B(t)}}\\
    -\sqrt{\frac{B(t)}{A(t)}}\Big(1-\frac{|x|}{\sqrt{x^2-\frac{1}{B(t)}}}\Big),              & |x|\geq \frac{1}{\sqrt{B(t)}}.
\end{cases}
\eea
This is the driving potential that can generate classical elliptical potential with sharp boundary in phase space.

Note that the value of $V(x,t)$ is divergent at $x=\pm {1}/{\sqrt{B(t)}}$ according to Eq.~(\ref{eq-ellipVxt-sm}). In fact, the Fourier coefficient Eq.~(\ref{eq-fFT}) is always divergent for $k\rightarrow \infty$ in the quantum regime $\lambda>0$. For physical result, we can add an exponentially suppressing factor to the Fourier coefficient: 
\bea
f_T(k_x.k_p)\rightarrow f_T(k_x.k_p)e^{-\frac{1}{2}\sigma^2 (k_x^2+k_p^2)}\ \  \mathrm{with}\ \  \sigma > \sqrt{{\lambda}/{2}}.
\eea
 According to the convolution theorem \cite{Arfken2005book}, we equivalently modify the Hamiltonian Q-function with a convolution operation, i.e., 
\bea\label{eq-HTsigma-sm}
H^{(T)}_{Q,\sigma}=\int\int dx'dp' g(x',p')H^{(T)}_Q(x-x',p-p')\ \ \mathrm{with}\ \ g(x,p)=\frac{1}{2\pi\sigma^2}e^{-\frac{x^2+p^2}{2\sigma^2}}.
\eea
The kernel function $g(x,p)$ smooths the sharp boundary of the elliptical well. Correspondingly, the driving potential is also modified with a convolution, i.e., 
\bea
V_\sigma(x,t)=\int\ dx'h(x')V(x-x',t)\ \ \mathrm{with}\ \ h(x)=\frac{1}{\sqrt{\pi(2\sigma^2-\lambda)}}e^{-\frac{x^2}{2\sigma^2-\lambda}}.
\eea
%
We point out that, in the Fock representation, the exponentially suppression factor $e^{-\frac{\lambda}{4}(k^2_x+k^2_p)}$ in Eq.~(\ref{eq-sm-nem})  cancels the same exponentially increasing factor in Eq.~(\ref{eq-sm-fnm}).   As a result, a sharp well with boundary narrower than quantum fluctuations ($\sigma<\sqrt{\lambda/2}$) does exist. However, this scenario cannot be realised by our present method.


\section{~ Moir\'e superlattice} \label{app-VIII}

In this section,  we discuss how to engineer a Moir\'e superlattice in phase space that is formed by two honeycomb phase space lattices overlaid with a relative twist angle and confined in a finite region with radius $R$,
\bea
H^{(T)}_Q(x,p)=
\begin{cases}
    H_{\theta=0}+H_{\theta=\theta_0},&  \sqrt{x^2+p^2}\leq R\\
    0,              & \sqrt{x^2+p^2}>R.
\end{cases}
\eea
Here, $H_\theta(x,p)\equiv-\prod_{n=1}^3\sin^2\big[\frac{1}{2}\tb v_n\cdot\tb z(\theta)\big]+\frac{3}{32}
$ is the honeycomb lattice in phase space \cite{Guo2022prb}.
We have defined the   vector ${\tb z}(\theta)\equiv (x\cos\theta+p\sin\theta,-x\sin\theta+p\cos\theta)$, and three ancillary vectors 
\bea
\tb v_1=(\frac{2\sqrt{3}}{3},0), \ \ \tb v_2=(-\frac{\sqrt{3}}{3},1), \ \ \tb v_3=(-\frac{\sqrt{3}}{3},-1). 
\eea
In Fig.~3(a) in the main text, we plot the resulting Moir\'e superlattice Q-function with twisted angle $\theta_0=10^\circ$. 

We calculate and plot the NcFT coefficient $f_T(k_x,k_p)$ in Fig.~3(b) in the main text, which composes of discrete peaks reflecting the discrete translational symmetry of target Hamiltonian in phase space. The centers of these peaks take place at $(k^q_x=k_q\cos\tau_q,k^q_p=k_q\sin\tau_q)$ where $\tau_q=q{\pi}/{6},\ q{\pi}/{6}+\theta_0$ and $k_q={2}/{\sqrt{3}},\ 2,\ {4}/{\sqrt{3}}$ with $q\in \mathbb{Z}$. The finite width of peaks comes from the boundary condition of Moir\'e superlattice. In fact, we have the analytical expression from Eq.~(\ref{eq-fFT}) 
\bea\label{eq-fTkxkpq}
f_T(k_x,k_p)=\sum_{q}\frac{A_q }{2\pi }\frac{J_1\Big(R\sqrt{(k_x-k^q_x)^2+(k_p-k^q_p)^2}\Big)}{R^{-1}\sqrt{(k_x-k^q_x)^2+(k_p-k^q_p)^2}},\ \  
\eea
where $A_q={\pi}/{16},\ -{\pi}/{16},\ {\pi}/{32}$ for $k_q={2}{\sqrt{3}}, \ 2,\  {4}{\sqrt{3}}$ respectively. 
Due to the long-distance asymptotic behavior of Bessel function
\bea
J_1(k)\approx\sqrt{\frac{2}{\pi k}}\cos(k-\frac{\pi}{4})\ \ \mathrm{ for}\ \  |k|\gg 1, 
\eea
we have added an exponentially suppressing factor $e^{-\frac{\lambda}{4} (k_x^2+k_p^2)}$ to obtain convergent NcFT coefficient. 
As a result, the Hamiltonian Q-function smoothed with a convolution kernel function $g(x,p)=\frac{1}{\pi\lambda}e^{-\frac{x^2+p^2}{\lambda}}$, cf. Eq.~(\ref{eq-HTsigma-sm}).

\section{~Hamiltonian operator and Q-function symmetry}\label{app-IX}

In this section, we prove that the Hamiltonian operator and its Q-function has the same symmetry in phase space. We first discuss  the Hamiltonian operator in the discrete rotational operation $\hat{R}_q\equiv e^{i\hat{a}^\dagger\hat{a}\frac{2\pi}{q}}$.
From the Fourier form of Hamiltonian operator Eq.~(\ref{eq-HTxp}), we have  
\bea\label{eq-HTxp-Rq}
\hat{R}_q\hat{H}(\hat{x},\hat{p})\hat{R}_q^\dagger
&=&
\frac{\beta}{2\pi}\int \int dk_x dk_pf_T(k_x,k_p)\hat{R}_q\exp[i(k_x\hat{x}+k_p\hat{p})]\hat{R}_q^\dagger\nl
&=&
\frac{\beta}{2\pi}\int \int dk_x dk_pf_T(k_x,k_p)\exp\Big[i\big([k_x\cos(\frac{2\pi}{q})-k_p\sin(\frac{2\pi}{q})]\hat{x}+[k_x\sin(\frac{2\pi}{q})+k_p\cos(\frac{2\pi}{q})]\hat{p}\big)\Big]\nl
&=&
\frac{\beta}{2\pi}\int \int dk'_x dk'_pf_T(k'_x,k'_p)\exp[i(k'_x\hat{x}+k'_p\hat{p})].
\eea
Here, we have used the property
\begin{eqnarray}\label{}
\left\{
\begin{array}{lll}
\hat{R}_q\hat{x}\hat{R}_q^\dagger=\hat{x}\cos(\frac{2\pi}{q})+\hat{p}\sin(\frac{2\pi}{q})\\
\hat{R}_q\hat{p}\hat{R}_q^\dagger=-\hat{x}\sin(\frac{2\pi}{q})+\hat{p}\cos(\frac{2\pi}{q}),\\
\end{array}
\right.
\end{eqnarray}
and transformed the integral coordinates by
\begin{eqnarray}\label{eq-kxkp-sm}
\left\{
\begin{array}{lll}
k'_x=k_x\cos(\frac{2\pi}{q})-k_p\sin(\frac{2\pi}{q})\\
k'_p=k_x\sin(\frac{2\pi}{q})+k_p\cos(\frac{2\pi}{q})\\
\end{array}
\right.
\end{eqnarray}
with the property $dk_xdk_p=dk'_xdk'_p$ as the transformation is orthogonal. 

Next, we will prove the NcFT coefficient of rotational lattice satisfies $f(k'_x,k'_p)=f(k_x,k_p)$. According to (\ref{eq-fFT}), we have
\bea\label{}
f_T(k'_x,k'_p)&=&\frac{e^{\frac{\lambda}{4}(k'^2_x+k'^2_p)}}{2\pi\beta}\int\int dxdp H^{(T)}_Q(x,p) e^{-i(k'_xx+k'_pp)}\nl
&=&\frac{e^{\frac{\lambda}{4}(k^2_x+k^2_p)}}{2\pi\beta}\int\int dx'dp' H^{(T)}_Q(x',p') e^{-i(k_xx'+k_pp')},
\eea
where we have used transformation Eq.~(\ref{eq-kxkp-sm}) and made the orthogonal transformation in the phase space plane
\begin{eqnarray}\label{}
\left\{
\begin{array}{lll}
x'=x\cos(\frac{2\pi}{q})+p\sin(\frac{2\pi}{q})\\
p'=-x\sin(\frac{2\pi}{q})+p\cos(\frac{2\pi}{q}).\\
\end{array}
\right.
\end{eqnarray}
Therefore, if the Hamiltonian Q-function satisfied the discrete rotational symmetry $H^{(T)}_Q(x',p')=H^{(T)}_Q(x,p)$, the NcFT coefficient of rotational lattice satisfies $f(k_x,k_p)=f(k'_x,k'_p)$. Then, comparing Eq.~(\ref{eq-HTxp}) and Eq.~(\ref{eq-HTxp-Rq}), we have $\hat{R}_q\hat{H}(\hat{x},\hat{p})\hat{R}_q^\dagger=\hat{H}(\hat{x},\hat{p})$ and vice versa .

The same discussion can also be applied for the translational symmetry described by the displacement operator 
\bea\label{eq-HTxp-D}
\hat{D}_{\alpha_0} \hat{H}(\hat{x},\hat{p})\hat{D}_{\alpha_0}^\dagger
&=&
\frac{\beta}{2\pi}\int \int dk_x dk_pf_T(k_x,k_p)\hat{D}_{\alpha_0}\exp[i(k_x\hat{x}+k_p\hat{p})]\hat{D}_{\alpha_0}^\dagger\nl
&=&
\frac{\beta}{2\pi}\int \int dk_x dk_pf_T(k_x,k_p)e^{-i(k_xx_0+k_pp_0)}\exp[i(k_x\hat{x}+k_p\hat{p})],
\eea
where we have used the property $\hat{D}_{\alpha_0} \hat{a}\hat{D}_{\alpha_0}^\dagger=\hat{a}-\alpha_0$ and thus 
\begin{eqnarray}\label{}
\left\{
\begin{array}{lll}
\hat{D}_{\alpha_0} \hat{x}\hat{D}_{\alpha_0}^\dagger=\hat{x}-x_0\  \mathrm{with} \ x_0=\langle \alpha_0|\hat{x}|\alpha_0\rangle\\
\hat{D}_{\alpha_0} \hat{p}\hat{D}_{\alpha_0}^\dagger=\hat{p}-p_0\  \mathrm{with} \ p_0=\langle \alpha_0|\hat{p}|\alpha_0\rangle.\\
\end{array}
\right.
\end{eqnarray}
From Eq.~(\ref{eq-fFT}), we have
\bea\label{}
f_T(k_x,k_p)e^{-i(k_xx_0+k_pp_0)}&=&\frac{e^{\frac{\lambda}{4}(k^2_x+k^2_p)}}{2\pi\beta}\int\int dxdp H^{(T)}_Q(x,p) e^{-i[k_x(x+x_0)+k_p(p+p_0)}\nl
&=&\frac{e^{\frac{\lambda}{4}(k^2_x+k^2_p)}}{2\pi\beta}\int\int dxdp H^{(T)}_Q(x+x_0,p+p_0) e^{-i(k_xx+k_pp)}.
\eea
Therefore, if the Hamiltonian Q-function satisfied the translational symmetry $H^{(T)}_Q(x+x_0,p+p_0)=H^{(T)}_Q(x,p)$, the NcFT coefficient of rotational lattice satisfies $f_T(k_x,k_p)e^{-i(k_xx_0+k_pp_0)}=f(k_x,k_p)$. Then, comparing Eq.~(\ref{eq-HTxp}) and Eq.~(\ref{eq-HTxp-D}), we have $\hat{D}_{\alpha_0} \hat{H}(\hat{x},\hat{p})\hat{D}_{\alpha_0}^\dagger=\hat{H}(\hat{x},\hat{p})$ and vice versa .

The same discussion can also be applied for other phase-space symmetries like the mirror symmetry, i.e.,  if the Hamiltonian Q-function satisfies $H^{(T)}_Q(x,p)=H^{(T)}_Q(\pm x,\pm p)$, the Hamiltonian operator has $ \hat{H}(\pm\hat{x},\pm\hat{p})=\hat{H}(\pm\hat{x},\pm\hat{p})$ and vice versa.

Lastly, we point out that the conclusion is also true for the smoothed Hamiltonian Q-function 
 with a convolution operation, i.e., 
\bea
H^{(T)}_{Q,\sigma}=\int\int dx'dp' g(x',p')H^{(T)}_Q(x-x',p-p')
\eea
 as long as the kernel function is rotationally symmetric, e.g., the standard Gaussian kernel $g(x,p)=\frac{1}{2\pi\sigma^2}\exp(-\frac{x^2+p^2}{2\sigma^2})$.

\twocolumngrid

\end{document}